\def\beq{\begin{equation}}
\def\eeq{\end{equation}}
\def\doeack{This work was supported in part by the Department of 
Energy, Nuclear Physics Division, under contract DE-FG02-86ER40286}
\documentclass{article}
\begin{document}
\title{Relativistic Quantum Mechanics -
Particle Production and Cluster Properties}
\author{W. N. Polyzou\thanks{\doeack}}
\maketitle
\begin{abstract}

This paper constructs relativistic quantum mechanical models of
particles satisfying cluster properties and the spectral condition which
do not conserve particle number.  The treatment of particle production
is limited to systems with a bounded number of
bare-particle degrees of freedom.  The focus of this paper is about the
realization of cluster properties in these theories.

\end{abstract}
\bigskip
\section{Introduction}

The purpose of this paper is to formulate a class of quantum theories
of interacting particles with the following properties.  They are
Poincar\'e invariant, they satisfy cluster properties, the
four-momentum operator has a spectrum supported in the future-pointing
light cone, and they allow particle production.  These theories are
applicable to problems in strong interaction physics where
relativistic invariance is an important symmetry.  Cluster properties
can be used to systematically build many-particle models from 
few-body models that are constrained by experiment.  Relativistic 
quantum theories with cluster properties are essential to the relevance 
of the few-body program at accelerators such as TJNAF.

The formulation of Poincar\'e invariant quantum theories satisfying
cluster properties for systems of a fixed number of particles has
been discussed in refs. \cite{sokolov}\cite{fcwp}
\cite{bkwp}\cite{wkwp}\cite{wp}.  This paper discusses the modifications 
to the fixed-$N$ construction needed to extend that construction to 
treat a class of models with particle production.   The theories 
discussed in this paper are limited to systems with a finite number
of bare-particle degrees of freedom.  A complete treatment of particle 
production, with no restrictions on the number of bare-particle degrees 
of freedom, is beyond the scope of this paper.
 
The physical properties mentioned in the first paragraph are the
minimal physical requirements for a realistic description of systems
strongly interacting particles.  The physical motivation for each of
these requirements is discussed below.

Poincar\'e invariance is the requirement that the group of continuous
Poincar\'e transformations is a symmetry of the theory.  In 1939
Wigner \cite{wigner} showed that this is equivalent to the existence
of a unitary representation of the Poincar\'e group on the Hilbert
space of the quantum theory.

Poincar\'e invariance is essential for a consistent interpretation of
any reaction with strong binding or particle production.  For
reactions where the initial and final states have different inertial
masses, momentum conservation cannot be simultaneously satisfied in
the laboratory and center of momentum frames in a Galilean invariant
quantum theory.

Cluster properties require that isolated subsystems have the same
properties as the system.  They relate interactions in subsystem
Hamiltonians to interactions in the system Hamiltonian.  Cluster
properties provide the connection between the few- and many-body
problem and the justification for experiments that are performed on
isolated targets at modern accelerators.

The spectral condition is essential for the stability of matter.
The mathematical requirement is that the eigenvalue spectrum of 
Hamiltonian is bounded from below.

Theories must be able to model reactions that change particle number.
These reactions are observed in experiments at almost all modern
accelerators.

While the above discussion makes a case that the physical
constraints discussed above are essential requirements for any
reasonable quantum theory of strongly interacting particles, it is
surprisingly difficult to formulate mathematically well-defined
theories that are consistent with all of these properties.

Even in quantum field theory, the problem of identifying the physical
Hilbert space, ${\cal H}$, and finding a set of ten self-adjoint
operators on this space that satisfy the Poincar\'e commutation
relations is an unsolved problem, except for the case of free quantum
fields.  These are the minimal requirements for realizing the
Poincar\'e symmetry in a quantum theory.

Some of the difficulties in formulating theories that 
are consistent with these physical constraints are discussed below.

The Poincar\'e group provides an infinite number of independent paths
to the future and each path involves the dynamics.  If one starts with
a given state and transforms it to a future time using different
combinations of Poincar\'e transformations, consistency requires that
the resulting states are identical.  For example, time
evolution can be expressed in terms of rotationless Lorentz
transformations and spatial translations.  Consistency of the 
quantum initial value problem requires that
if there are interactions in the Hamiltonian then there must be
interactions in the infinitesimal generators of rotationless Lorentz
transformations and/or spatial translations.  This is a
consequence of the commutation relation
\beq
[{P}_i,{K}_j ] = i \delta_{ij} {H},
\label{eq:AA}
\eeq
which relates the Hamiltonian, $H$, to the linear momentum generators,
$\vec{P}$, and generators of rotationless Lorentz transformations,
$\vec{K}$.  The Poincar\'e commutation relations impose non-linear
constraints on these interactions.

Cluster properties impose independent non-linear constraints on the
interactions.  To see this note that in the three-body problem,
cluster properties fix the two-body interactions in each of the
Poincar\'e generators up to an overall three-body interaction.
However, because interactions involving different pairs of particles
appear in more than one generator, the operators obtained by adding
the required two-body interactions to the non-interacting generators
fail to satisfy the commutation relations without additional
three-body interactions.  For example, if the generators of
rotationless Lorentz transformations, $\vec{K}$, have interactions
between particles 1 and 2 and the Hamiltonian has interactions between
particles 2 and 3, then the commutator, $[H,\vec{K}] = i \vec{P}$ will
have three-body interactions involving particles 1, 2, and 3 unless $H$
and/or $\vec{K}$ have three-body interactions that are designed to cancel the
three-body operators generated by the commutator. 

 While the spectral condition is not difficult to satisfy, negative
energy states have historically appeared when classical relativistic
field theories, like the Klein-Gordon-Schr\"odinger and Dirac
equations, are treated as quantum mechanical equations.  The negative 
energy eigenstates of the Hamiltonian disappear when these equations 
are properly treated as equations for quantum fields.

Particle production requires a more critical analysis of cluster
properties.  For theories with a fixed number, $N$, of particles there
is an ordering on particle number and cluster properties define the
relationship between the interactions in the $K<N$-body Poincar\'e
generators and the $N$-body Poincar\'e generators.  This leads to 
important relations between the dynamics of the system and its 
proper subsystems.  These relations provide the justification for 
both theory and experiment on few-body systems.   

The problem with formulating a useful cluster condition in theories
with particle production is the absence of a few-body 
problem that puts  useful constraints on the many-body dynamics.
Specifically,  in theories with particle production, states with
a few physical particles generally involve an infinite number of bare
particles. 

The difficulties with formulating quantum theories with an infinite
number of degrees of freedom are well known \cite{haag}\cite{fcrh}
\cite{araki}.  These difficulties are distinct from the specific 
problems that arise from particle production.  In this paper these
problems are deliberately separated by restricting considerations to a
class of theories with a finite number of bare-particle degrees 
of freedom.  This is achieved using conservation laws that limit the number of
bare-particle degrees of freedom.  It is possible to formulate cluster
properties in these theories without having to confront the specific
problems that arise due to the infinite number of degrees of freedom.
  
The class of models considered in this paper are designed to
complement models based on formal quantum field theory.  Quantum
mechanical models of interacting particles have the advantage that
(for systems of strongly interacting particles) they are
mathematically well-defined and can in principle be solved using
convergent algorithms.  It is for this reason that quantum theories of
particles are often used to model few-body reactions involving
composite systems or scattering from composite targets.  Some recent
applications can be found in \cite{fuda2}\cite{gianni}
\cite{klink}\cite{glockle}.  Extending these theories to
Poincar\'e invariant theories with cluster properties 
that allow particle production provide a
more robust class of models.

The next section discusses the assumptions that are used to limit the
number of bare-particle degrees of freedom.  The structure of the
model Hilbert space is given in the following section.  It differs
from the Hilbert space for a system of $N$ particles in how it factors
into subsystem spaces.  This factorization and some of its properties
are discussed in section four.  In section five the cluster property
is formulated in a manner that is consistent with the modified
factorization into subsystems.  The formulation of scattering theory
for reactions that do not conserve particle number is discussed in
section six.  Modification of the $C^*$ algebra of asymptotic
constants, which is a central element of the construction of a
dynamics satisfying cluster properties in ref. \cite{wp}, is 
discussed in section seven.
The unitary elements of this algebra preserve the scattering
observables, and can be used to restore cluster properties.  The
modifications to the general construction in \cite{wp} to treat
variable numbers of particles are summarized in section eight.  Rather
than give a systematic description of the general construction, as was
done in ref. \cite{wp}, the essential elements of the general
construction are illustrated in sections 9-11 using a non-trivial
example.

\section{\bf Motivation and Assumptions:}

General features of the class of theories studied in this paper are
motivated by comparing theories of a fixed number of particles 
to theories that change particle number.  The construction in this
paper extends the general fixed-$N$ construction in \cite{wp}.  In 
all that follows we use notation from \cite{wp}.

Consider a relativistic theory of $N$-interacting particles following
the construction of \cite{wp}.  Relativistic invariance is realized by
a dynamical unitary representation ${U}[\Lambda ,Y]$ of inhomogeneous
$SL(2,C)$ $(ISL(2,C))$ on the $N$-particle Hilbert space, ${\cal H}$.
$ISL(2,C)$ is the covering group for the Poincar\'e group; 
it is used because the relevant representations are single valued 
and computations are easier using $2 \times 2$ matrices.

The $SL(2,C)$ matrix  
$\Lambda$ is related to a finite Lorentz transformation 
$\Lambda^{\mu}{}_{\nu}$ by
\beq
\Lambda^{\mu}{}_{\nu} = {1 \over 2} \mbox{Tr} \left ( \sigma_{\mu}\Lambda 
\sigma_{\nu} \Lambda^{\dagger}\right ) ,  
\label{eq:BA}
\eeq
and the $2 \times 2$ Hermitian matrix, $Y$, parameterizes a space-time 
translation $y^{\mu}$ by 
\beq
Y = y^{\mu} \sigma_{\mu} \qquad y^{\mu} = 
{1 \over 2} \mbox{Tr}(Y \sigma_{\mu}) .
\label{eq:BB}
\eeq
The group product is
\beq
(\Lambda_2,Y_2) (\Lambda_1,Y_1)= 
(\Lambda_2\Lambda_1,\Lambda_2 Y_1 \Lambda_2^{\dagger} +Y_2).
\label{eq:BC}
\eeq  
The resulting ${U}[\Lambda,
Y]$ satisfies cluster properties and the spectrum of the Hamiltonian,
${H}$, is bounded from below. 

Assume that in this model some
isolated subsystems can form bound states.  Then cluster properties imply
that the isolated bound subsystems have the same Poincar\'e
transformation properties as elementary particles with the same mass
and spin.  With respect to their Poincar\'e transformation properties,
there is no distinction between elementary and composite particles.

Treating the asymptotically stable subsystems as physical particles,
the relativistic theory described in \cite{wp} can be interpreted as a
theory of fixed number of bare particles with a variable number of
physical particles.  The physical particles in the above sense are
needed to formulate scattering asymptotic conditions and cluster
properties.

This can be compared to local quantum field theory, where physical
particles, defined as discrete eigenstates of the mass and spin, also
have a composite bare-particle content.  An important distinction is
that in local field theory the physical particles involve an infinite
number of bare-particle degrees of freedom, while in the relativistic
quantum mechanics case discussed above, the composite systems involve
a fixed finite number of bare-particle degrees of freedom.

In this paper the fixed-$N$ construction is generalized by replacing
the $N$-constituent particles a set of conserved additive quantum
numbers.  These quantum numbers have no physical interpretation; they
are introduced to provide a mechanism to control the number of degrees 
of freedom.  These quantum numbers are called charges and 
they are assumed to satisfy:

\begin{itemize} 

\item[a.] There are $K$ types of charge.

\item[b.] Charges can have only {\it non-negative} integer values.  

\item[c.] Each bare particle of the model has a set of charges
labeled by an $n$-tuple of integers $(n_1, \cdots, n_K)$ labeling 
the number of each of the $K$-types of charges.
 
\item[d.] The charge of a composite system is the sum of the charges of the 
constituents.

\item[e.] Each bare particle of the theory has as least one non-zero charge.

\item[f.] Interactions conserve all $K$ types of charge.

\end{itemize}

The charge of a bare particle is {\it minimal} if it cannot
be expressed as a sum of smaller charges corresponding to 
at least two bare particles.

The relativistic Lee model
\cite{lee}\cite{dormale}\cite{fcwp}\cite{fuda} provides a well-known
example of a theory with this structure.  The Lee model has three
types of bare particles which can be suggestively called a $\pi$, $N$,
and $\Delta$, with a vertex interaction $\pi + N\leftrightarrow
\Delta$. There are two conserved charges $(q_N,q_\pi)$ where the $\pi$
has charge $(0,1)$, the $N$ has charge $(1,0)$, and the $\Delta$ has
charge $(1,1)$.  In this model the charges of the $\pi$ and the $N$ are
minimal.  The charge of the $\Delta$ is not minimal because the
$\pi$-$N$ system has the same charge as the $\Delta$.
In this model the $\Delta$ is called a {\it composite} bare particle.
The vertex interaction conserves charge.   Many isobar models also
fall into this class.

Theories with an infinite number of degrees of freedom are obtained by
dropping the assumptions $b.)$ and $e.)$. For example, if a neutral
pion is assigned a charge zero and a neutron is assigned a charge equal to its
baryon number, each fixed-charge subspace of the Hilbert space has
subspaces with arbitrarily large numbers of pions and
neutron anti-neutron pairs.  This paper only considers theories where
conditions $b.)$ and $e.)$ are enforced.  With these restrictions 
it is possible to define a meaningful ``few-charge'' problem.
   
\section{Hilbert Space}

The Hilbert space ${\cal H}_{\{N\}}$, corresponding to the value
$\{N\}= (n_1 \cdots n_K)$ of the conserved charges is a
direct sum of tensor products of bare-particle Hilbert spaces,
\beq
{\cal H}_{\{N\}}:= \oplus_{i=1}^n (\otimes_{k=1}^{n_i} 
{\cal H}_{m_{ik} j_{ik}}) 
\label{eq:CA}
\eeq 
where ${\cal H}_{mj}$ is the mass $m$, spin $j$ irreducible
representation space of $ISL(2,C)$.  Each term of the direct sum has a
{\it different bare-particle content}, but the same value of total charge,
$\{N\}$.

In the Lee model example the Hilbert space,
\beq
{\cal H}_{\{1,1\} } = ({\cal H}_N \otimes {\cal H}_\pi) 
\oplus {\cal H}_{\Delta}
\label{eq:CB}
\eeq
is the direct sum of the two-particle $N-\pi$ space and the 
one-particle $\Delta$ space.    Note that including a bare $\Delta$ 
particle in the model does not imply that the $\Delta$ 
will exist as a 
stable physical particle. 

The irreducible representation spaces, ${\cal H}_{mj}$, of $ISL(2,C)$ 
are spaces of
square integrable functions of the eigenvalues of a maximal set of
commuting self-adjoint functions of the single particle generators.
In general this set includes the invariant mass and spin operators, and
four additional functions \cite{wp} of the $ISL(2,C)$ generators, 
denoted by
${F}^i$. A typical choice of 
the operators ${F}^i$ is the three components
of the linear momentum and the $\hat{z}$-component of the canonical spin.

Denoting the eigenvalues of ${F}^i$, $m$, $j^2$ by $f$, $m$,
and $j(j+1)$ gives basis vectors on each ${\cal H}_{mj}$ of the form
\beq
\vert f ;m,j \rangle
\label{eq:CC}
\eeq
with resolution of the identity and 
normalization given by 
\beq
I = \int \vert f ;m,j \rangle d\mu (f) \langle  f ;m,j \vert
\qquad
\langle f ; m,j  \vert f' ; m,j \rangle = \delta [f,f'].
\label{eq:CD}
\eeq
In this expression $\int d\mu (f)$ denotes an integral over the
continuous eigenvalues and a sum over the discrete eigenvalues of
${F}^i$.  Likewise, $\delta [f,f']$ indicates a product of Dirac
delta functions in the continuous variables and Kronecker delta
functions in the discrete variables.  
This basis of the single 
particle Hilbert space is called the $f$-basis.  

By assumption, $ISL(2,C)$ acts irreducibly
on this space.  In the $f$-basis the action of 
${U}[\Lambda ,Y]$ is given by:
\beq
{U}[\Lambda ,Y] \vert f ;m,j \rangle
= \int \vert f' ;m,j \rangle d \mu (f')
{\cal D}^{mj}_{f'f}[\Lambda ,Y]
\label{eq:CE}
\eeq
where 
\beq
{\cal D}^{mj}_{f'f}[\Lambda ,Y] := 
\langle f' ;m,j \vert {U}[\Lambda ,Y] \vert f ;m,j \rangle
\label{eq:CF}
\eeq
is the mass $m$ spin $j$ irreducible representation if $ISL(2,C)$ in
the $f$-basis.   What is relevant for this paper is that the 
irreducible representations 
${\cal D}^{mj}_{f'f}[\Lambda ,Y]$ are known for each value of
$m$ and $j$.   Explicit formulas for Poincar\'e ${\cal D}$-functions, 
${\cal D}^{mj}_{f'f}[\Lambda ,Y]$,
are given in \cite{wp2}\cite{bkwp}\cite{wp}.

The irreducible representation, ${U}_{ik}[\Lambda ,Y]$, of
$ISL(2,C)$ on each of the subspaces ${\cal H}_{m_{ik}j_{ik}}$
can be used to construct a natural {\it non-interacting} representation,
${U}_{0}[\Lambda ,Y]$, on ${\cal H}_{\{N\}}$ given by:
\beq
{U}_0 [\Lambda ,Y] = 
\sum_{i=1}^n (\otimes_{k=1}^{n_i} {U}_{ik} [\Lambda ,Y] ).
\label{eq:CG}
\eeq

The Hilbert space ${\cal H}_{\{N\}}$ has two natural bases.  
The first is the tensor product of single bare-particle basis vectors.
There are distinct basis functions corresponding to each 
orthogonal subspace in the direct sum (\ref{eq:CA}).

The second is a basis that transforms irreducibly with respect to
${U}_0 [\Lambda ,Y]$.  The irreducible basis is constructed, using the
$ISL(2,C)$ Clebsch-Gordan coefficients
\cite{wick}\cite{Joos}\cite{fc2}\cite{bkwp}\cite{wkwp}\cite{wp}, 
as a linear combination of
the tensor product of irreducible representations.  As in the case of
the tensor product basis, there is a distinct orthogonal subspace
corresponding to each term in the direct sum (\ref{eq:CA}).

The two types of basis vectors on ${\cal H}_{\{{\cal N}\}}$ are denoted by 
\beq
\vert \otimes f_i;j_i,m_i \rangle
\label{eq:CH}
\eeq
and 
\beq
\vert f, d; j,m \rangle
\label{eq:CI}
\eeq
respectively, where $d$ denotes a set of invariant degeneracy quantum
numbers.  The $m$ in (\ref{eq:CI}) is the invariant mass of the 
system of non-interacting bare particles in the tensor product.

The second basis transforms irreducibly with respect to 
${U}_0 [\Lambda ,Y]$:
\beq
{U}_0 [\Lambda ,Y] \vert f ,d ;m,j \rangle =
\int \vert f' ,d ;m,j \rangle d\mu (f') {\cal D}^{mj}_{f'f} [\Lambda, Y] ,
\label{eq:CJ}
\eeq
which has the same form as the transformation law for a single particle
of mass $m$ and spin $j$, while the basis (\ref{eq:CH}) transforms like
\beq
{U}_0 [\Lambda ,Y]  \vert \otimes f_i;j_i,m_i \rangle =
\int  \vert \otimes f_i';j_i,m_i \rangle 
\prod_l d\mu (f_l') {\cal D}^{m_lj_l'}_{f_l'f_l} [\Lambda, Y] .
\label{eq:CK}
\eeq

\section{Tensor Products/Factorization} 

In non-relativistic many-particle quantum mechanics the $N$-particle
Hilbert space can be decomposed into a tensor product of Hilbert
spaces with fewer particles.  Cluster properties lead to an asymptotic
factorization of interacting representation, ${U}[\Lambda ,Y]$, into a
tensor product of subsystems ${U}_{a_i} [\Lambda,Y]$'s that act on
each factor of the tensor product

For models with conserved charges $\{ N \}$ a similar, but slightly 
more complicated relationship exists.  To define this relationship begin
by labeling each charge.  In this paper the charges are initially 
treated as distinguishable.   Proper symmetry under exchange of identical 
particles can be restored after the Poincar\'e generators are constructed. 
 
Let ${\cal H}_{\{{\cal N}\}}$ be the Hilbert space for a given set of
charges.  Partitions $a$ of the labeled charges are identified with 
equivalence relations on the charges.  The $i$-th equivalence class, denoted by $a_i$,
called the $i$-th cluster of $a$.  The set of all partitions of
the charges is denoted by ${\cal P}_{\{\cal N\}}$.

To each partition $a$ of the conserved charges the Hilbert space
${\cal H}_{\{N\}}$ can be decomposed as an orthogonal direct sum of the form:
\beq
{\cal H}_{\{N\}} = {\cal H}_a \oplus {\cal H}^a
\label{eq:DA}
\eeq
where 
\beq
{\cal H}_a :=  \otimes_{i=1}^{n_a} {\cal H}_{\{N_{a_i}\}}
\label{eq:DB}
\eeq
is the tensor product of the subsystem Hilbert spaces associated with 
the charges in the $i$-th cluster of $a$ and
${\cal H}^a$ is the orthogonal complement of ${\cal H}_a$ in 
${\cal H}_{\{N\}}$. 

The residual space, ${\cal H}^a$, appears because for each 
partition, $a$, the Hilbert space ${\cal H}_{\{{\cal N}\}}$ may have 
a subspace with bare particles having non-zero 
charges in {\it different} clusters of the partition $a$.

In the case of the $\{1,1\}$ sector of the Lee 
model \cite{lee}\cite{dormale}\cite{fcwp}\cite{fuda}
the factorization (\ref{eq:DA}) has the form
\beq
{\cal H}_{\{1,1\}}= ({\cal H}_{N} \otimes {\cal H}_{\pi}) 
\oplus {\cal H}_{\Delta}
\label{eq:DF}
\eeq
where for $a=(N)(\pi)$, ${\cal H}_a ={\cal H}_{N} \otimes {\cal H}_{\pi}$
and ${\cal H}^a= {\cal H}_{\Delta}$.
The $\Delta$ subspace is unimportant for understanding clustering into an 
asymptotically separated $\pi$ and $N$.

The appearance and treatment of the residual space is the main
technical difference between models of a fixed number of particles and
models with production.

The partitions of the conserved charges into disjoint
equivalence classes has an obvious partial ordering given by: $ a
\supseteq b$ if and only if conserved charge labels in the same
equivalence class with respect to $b$ are in the same equivalence
class with respect to $a$.  This means that the clusters of $b$ 
are obtained by breaking up the clusters of $a$. 

Given the partial ordering on ${\cal P}_{\{N\}}$ it is possible to 
define M\"obius and Zeta functions \cite{kkwp}\cite{rota} for the 
partial ordering:
\beq
\zeta (a \supseteq b ) =
\left \{
\begin{array}{ll}
1 & a \supseteq b \\
0 & \mbox{otherwise} 
\end{array}
\right.
\label{eq:DC}
\eeq
and
\beq
\mu (a \supseteq b ) = \zeta^{-1} (a \supseteq b ) =
\left \{
\begin{array}{ll}
(-)^{n_a} \prod_{i=1}^{n_a} (-)^{n_{b_i}}(n_{b_i}-1)! & a \supseteq b \\
0 & \mbox{otherwise} 
\end{array}
\right.
\label{eq:DD}
\eeq
where $n_a$ is the number of clusters of the partition $a$ and $n_{b_i}$ is 
the number of clusters of $b$ in the $i$-th cluster of $a$.

It is a consequence of the definitions that the subspaces ${\cal H}_a$ and 
${\cal H}^a$ satisfy the relations: 
\beq
b \supset a \Rightarrow {\cal H}_{b} \supset 
{\cal H}_{a} \quad \mbox{and} \quad
{\cal H}^{a} \supset 
{\cal H}^{b} .
\label{eq:DE}
\eeq
This is equivalent to the observation that if a bare particle has
non-zero charges in two different clusters of a partition $a$, then
this is also true for any refinement of $a$.  This means that for the
purpose of studying cluster properties, the residual spaces ${\cal
H}^a$ can be ignored when making refinements of partitions.

\section{Cluster Properties}

In order to formulate cluster properties assume that it is possible to
find the dynamics
\beq
{U}_{a_i} [\Lambda ,Y]
\label{eq:EA}
\eeq
associated with the conserved charges in the $i$-th cluster of $a$.

The {\it interacting} representation ${U}[\Lambda ,Y]$ of the
system satisfies space-like cluster properties if for each partition
$a$ of the conserved charges the following strong limits
vanish:
\beq
\lim_{min (y_i-y_j)^2 \to \infty}  
\left ( {U}[\Lambda ,Y] - \otimes_{l=1}^{n_a} {U}_{a_l}[\Lambda ,Y]
\right )  \otimes_{m=1}^{n_a} {U}_{a_m}[I ,Y_m] \Pi_a = 0 ,
\label{eq:EB}
\eeq
where $\Pi_a$ is the orthogonal projection onto the subspace 
${\cal H}_a$ of ${\cal H}$. This projection is needed because  
$\otimes_{l=1}^{n_a} {U}_{a_l}[\Lambda ,Y]$ and 
$\otimes_{m=1}^{n_a} {U}_{a_m}[I ,Y_m]$ are only defined on 
${\cal H}_a$. For successive limits the projections should be 
on the largest subspace that allows the charges to be asymptotically 
separated.

Equation (\ref{eq:EB}) contains two conditions.  First, it requires that when
the interaction terms between particles with charges in different
clusters of $a$ are turned off, the projection of ${U}[\Lambda
,Y]$ on ${\cal H}_a$ becomes a tensor product of subsystem
representations.  This property is referred to the algebraic cluster
property.  This condition is non-trivial; when it fails either the 
cluster limit does not exists, or interactions between particles in the 
{\it same} cluster of $a$ vanish in the cluster limit \cite{bkwp}.

The second condition is that the interaction between particles with
charges in different clusters satisfy the short-range condition
specified above.  This can be reformulated as a ``Cook-like''
condition on the range of the residual interactions.  To see 
this let $[\Lambda
,Y]$ denote a fixed $ISL(2,C)$ transformation.  To formulate the
range condition assume
\beq
{U}[\Lambda ,Y] = e^{i {G} }
\label{eq:EC}
\eeq
where $G=G[\Lambda,Y]$, is a fixed linear combination
of the generators $ISL(2,C)$ on ${\cal H}$.

The limiting form required by cluster properties when the clusters of 
$a$ are asymptotically separated is
\beq
{U}_a[\Lambda ,Y] = e^{i (\sum{i} {G}_{a_i} )}=
e^{i {G}_{a} }.
\label{eq:ED}
\eeq
To formulate the cluster condition define the residual interaction by 
\beq
V^a = G - G_a .
\label{eq:EE}
\eeq
Consider
\beq
{F}(\alpha) := e^{i \alpha  {G}} e^{-i \alpha {G}_a} 
\label{eq:EF}
\eeq
where $e^{-i \alpha {G}_a}$ is extended to be the identity 
on ${\cal H}^a$. This satisfies the integral equation
\beq
{F}(\alpha)= I + i \int_0^\alpha {F}(\alpha') {V}^a (\alpha')
d\alpha'  
\label{eq:EG}
\eeq
where
\beq
{V}^a (\alpha ) := 
e^{i \alpha {G}_a} {V}^a e^{- i \alpha {G}_a} .
\label{eq:EH}
\eeq
The cluster condition is equivalent to 
\beq
\lim_{min (y_i-y_j)^2 \to \infty}
\Vert [{F}(1) - {I} ] \otimes_{l=1}^{n_a}{U}_{a_l} [I,Y_l] 
{\Pi}_a \vert \xi \rangle \Vert = 0.
\label{eq:EI}
\eeq
This limit is bounded by 
\beq
\lim_{min (y_i-y_j)^2 \to \infty} \int_0^1 \Vert 
{V}^a (\alpha') 
\otimes_{l=1}^{n_a}{U}_{a_l} [I,Y_l] 
{\Pi}_a  \vert \xi \rangle  \Vert d\alpha'.  
\label{eq:EJ}
\eeq
The integrand is uniformly bounded in $\alpha'$ by 
\beq
\Vert 
{V}^a \Vert  \Vert {\Pi}_a  \vert \xi \rangle  \Vert 
< \infty 
\label{eq:EK}
\eeq
and each term in the integrand has the limit
\beq
\lim_{min (y'_i-y'_j)^2 \to \infty} \Vert 
{V}^a 
\otimes_{l=1}^{n_a}{U}_{a_l} [I,Y'_l] 
{\Pi}_a  \vert \xi' \rangle  \Vert =0   
\label{eq:EL}
\eeq
where 
\beq
Y'_l = \Lambda Y_l \Lambda^{\dagger}+Y
\label{eq:EM}
\eeq
and 
\beq
\vert \xi' \rangle = e^{- i \alpha {G}_a}
\vert \xi \rangle .
\label{eq:EN}
\eeq
In this expression $\Lambda,Y$ is the $ISL(2,C)$ transformation 
defined by ${U}[\Lambda,Y] = e^{-i \alpha {G}}$.

It follows from the Lebesgue dominated 
convergence theorem \cite{rudin} that the 
cluster limit vanishes provided condition 
(\ref{eq:EL}) holds for all $\vert \xi' \rangle$ and
all asymptotic space-like separations, $(y_i'-y_j')^2 \to \infty$.  
This is the desired
``Cook-like'' condition on the range of the inter-cluster interaction,
$V^a$.  This is analogous to the cluster condition in non-relativistic quantum 
models. 

The difficult aspect of the cluster problem in relativistic quantum
mechanics is to construct ${U}[\Lambda ,Y]$ so it satisfies algebraic
cluster properties.  In this case algebraic cluster properties mean
that
\beq
{U}[\Lambda ,Y] \to  {U}_{a}[\Lambda ,Y]
\label{eq:EO}
\eeq
when the interactions between particles in different clusters are 
turned off, where
\beq
U_a [\Lambda ,Y] := 
\left (
\begin{array}{cc}
\otimes_{l=1}^{n_a} {U}_{a_i}[\Lambda ,Y] &0 \\
0& I 
\end{array}
\right ).     
\label{eq:EP}
\eeq
on ${\cal H}={\cal H}_a\oplus {\cal H}^a$.
Extending $\otimes_{l=1}^{n_a} {U}_{a_i}[\Lambda ,Y]$ to all of 
${\cal H}$ by extending it as the identity on the subspace 
${\cal H}^a$ is one of the modifications introduced because 
of the factorization (\ref{eq:DA}).  
The identity term is consistent with setting the corresponding generators to 
zero.  This choice does not effect the cluster condition because 
the identity term is eliminated by the projector $\Pi_a$.

This formulation of cluster properties has the property that if the
system is further sub-divided by $b \subset a$ then $\Pi_a \Pi_b =
\Pi_b$.  This is because all of the refinements of $a$ are defined 
on ${\cal H}_a$.  This property of the model Hilbert space ensures that system
can continue to be subdivided until all that remains is a system of
bare particles with minimal charges.  

\section{Scattering} 

The formulation of scattering theory with particle production 
is identical to the two-Hilbert
space formulation used in \cite{fcwp}\cite{wp} for 
fixed numbers of particles.

To formulate the scattering theory assume that the dynamical
representation ${U}[\Lambda ,Y]$ of $ISL(2,C)$ on ${\cal H}$ is given.
Assume that the representation
${U}[\Lambda ,Y]$ has the following properties, which are
consistent with the fixed $N$ case.

\begin{itemize}

\item[a.] There are simultaneous eigenstates of ${M}, {j}, {F}^i$
with positive discrete mass eigenvalues that transform irreducibly with 
respect to ${U} [\Lambda ,Y]$.

\item[b.] There are simultaneous eigenstates of ${M}, {j}, {F}^i$
with positive eigenvalues in the absolutely continuous spectrum
of ${M}$.  These satisfy scattering asymptotic conditions.

\item[c.] The bound and scattering eigenstates are complete on the model
Hilbert space, with the incoming and outgoing wave scattering states 
each spanning the orthogonal complement of the subspace spanned by the
bound states. 

\end{itemize} 

A bound state is a simultaneous eigenstate of ${F}_i, {M}$ and ${j}^2$: 
\beq
\vert f ;m,j \rangle 
\label{eq:FA}
\eeq
with discrete mass eigenvalue, $m$.
It transforms irreducibly under the action of the {\it dynamical} 
representation of $ISL(2,C)$:
\beq
{U} [\Lambda ,Y] \vert f ;m,j \rangle =
\int d \mu (f') \vert f';m,j \rangle
{\cal D}^{mj}_{f'f}[\Lambda ,Y]. 
\label{eq:FB}
\eeq
The function ${\cal D}^{mj}_{f'f}[\Lambda ,Y]$ is the known 
mass-$m$ and spin-$j$ irreducible representation of $ISL(2,C)$. 

Normalizable eigenstates of physical mass and spin 
can be expressed in the form
\beq
\vert \psi \rangle = \int \vert f ;m,j \rangle d \mu (f) \chi (f) 
\label{eq:FC}
\eeq
for square integrable functions $\chi (f)$. 

Each irreducible bound subspace defines a bound-state channel, $\alpha$. 
The channel Hilbert space, 
${\cal H}_\alpha $, is the space of square integrable functions,
$\chi (f)$, over the joint spectrum, $\sigma ({F})$, of the commuting 
operators $F^i$.  
Equation (\ref{eq:FC})
can be interpreted as a 
mapping ${\Phi}_{\alpha}$ from the bound channel Hilbert space 
${\cal H}_{\alpha}$ to the physical Hilbert space ${\cal H}$:
\beq
\vert \psi \rangle = {\Phi}_{\alpha} \vert \chi \rangle.
\label{eq:FD}
\eeq
This can be done for each bound channel.  Note that in general 
$\vert \psi \rangle$ has components in all bare-particle sectors of ${\cal H}$.

In this notation equation (\ref{eq:FB}) can be expressed in the form
\beq
{U} [\Lambda, Y] {\Phi}_{\alpha} =
{\Phi}_{\alpha} {U}_{\alpha} [\Lambda, Y] 
\label{eq:FE}
\eeq
where ${U}_{\alpha} [\Lambda, Y]$ is the irreducible 
unitary representation of $ISL(2,C)$ with kernel
${\cal D}^{mj}_{f'f}[\Lambda ,Y]$. 

Individual subsystem bound states are used to formulate the asymptotic 
condition for multi-particle scattering channels. 

For a partition $a$ of conserved charges the physical Hilbert space has the 
factorization ${\cal H}_{\{N\}}= {\cal H}_a\oplus {\cal H}^a$
where
\beq
{\cal H}_a = \otimes_{a_i} {\cal H}_{\{N_{a_i}\}}.
\label{eq:FF}
\eeq
   
Assume that there is a subsystem dynamics,
\beq
{U}_{a_i} [\Lambda ,Y]: {\cal H}_{\{N_{a_i}\}} \to {\cal H}_{\{N_{a_i}\}}
\label{eq:FG}
\eeq
for the charges in the $i$-th cluster of the partition $a$.  
There is a scattering
channel $\alpha$ associated with the partition $a$ if there is a bound 
channel for each of the subsystem ${U}_{a_i} [\Lambda ,Y]$'s.

Following (\ref{eq:FE}), for each 
bound subsystem there is an asymptotic Hilbert space 
${\cal H}_{\alpha_i}$ and an injection operator $\Phi_{\alpha_i}$:
\beq
\Phi_{\alpha_i}: {\cal H}_{\alpha_i} \to {\cal H}_{\{N_{a_i}\}}
\label{eq:FH}
\eeq
with the property 
\beq
{U}_{a_i} [\Lambda, Y] {\Phi}_{\alpha_i} = 
{\Phi}_{\alpha_i}{U}_{\alpha_i} [\Lambda, Y] .
\label{eq:FI}
\eeq
Define the channel Hilbert space ${\cal H}_{\alpha}$ and the channel 
injection operator ${\Phi}_{\alpha}:{\cal H}_{\alpha} \to
{\cal H}_{a} \subset {\cal H} $ by:
\beq
{\cal H}_{\alpha} := \otimes_{i=1}^{n_a} {\cal H}_{\alpha_i} 
\label{eq:FJ}
\eeq
\beq
{\Phi}_{\alpha}:= \otimes_{i=1}^{n_a} {\Phi}_{\alpha_i}.
\label{eq:FK}
\eeq

Define
\beq
{U}_a [\Lambda ,Y] :=
\left (
\begin{array}{cc}
\otimes_{i=1}^{n_a} {U}_{a_i} [\Lambda ,Y]  & 0 \\ 
0 & {I} \\ 
\end{array} 
\right )
\label{eq:FL}
\eeq
where the ${I}$ acts on the residual subspace ${\cal H}^a$.
Also define 
\beq
{U}_{\alpha}[\Lambda ,Y]:{\cal H}_{\alpha} \to {\cal H}_{\alpha} 
\label{eq:FM}
\eeq
by
\beq
{U}_{\alpha}[\Lambda ,Y] := \otimes_{i=1}^{n_a} 
{U}_{\alpha_i} [\Lambda ,Y] .
\label{eq:FN}
\eeq
With these definitions relations (\ref{eq:FI}) can be compactly expressed 
in the form
\beq
{U}_a [\Lambda ,Y]{\Phi}_{\alpha} =  
{\Phi}_{\alpha} {U}_{\alpha} [\Lambda ,Y].
\label{eq:FO}
\eeq 

Following the case of a fixed number of particles, a scattering state
is a solution 
\beq
\vert \psi_{\alpha}^{\pm}(t) \rangle =
{U}[I,T] \vert \psi_{\alpha}^{\pm}(0) \rangle
\label{eq:FP}
\eeq
of the time-dependent Schr\"odinger equation satisfying the 
asymptotic condition
\beq
\lim_{t \to \pm \infty} \Vert \psi_{\alpha}^{\pm}(t) \rangle
- {U}_a [I,T] {\Phi}_{\alpha} \vert \chi_{\alpha} \rangle \Vert 
=0
\label{eq:FQ}
\eeq 
for all $\vert \chi_{\alpha} \rangle \in {\cal H}_{\alpha}$.  Using the 
intertwining relations (\ref{eq:FO}) this can be expressed as
\beq
\lim_{t \to \pm \infty} \Vert \psi_{\alpha}^{\pm}(0) \rangle
- {U}^{\dagger} [I,T] {\Phi}_{\alpha} {U}_{\alpha} [I,T]
\vert \chi_{\alpha} \rangle \Vert 
=0.
\label{eq:FR}
\eeq 
As in the fixed number of particle case, this is automatically satisfied 
for the system bound states.  When this limit exists, channel wave operators 
are defined by the strong limits 
\beq
{\Omega}_{\alpha}^{\pm} = \lim_{t \to \pm \infty} 
{U}^{\dagger} [I,T] {\Phi}_{\alpha} {U}_{\alpha} [I,T] .
\label{eq:FS}
\eeq
Cook's condition \cite{simon} for scattering provides a 
sufficient condition for the existence of the channel wave operators.
\beq
\int_c^{\pm \infty} \Vert {V}_{\alpha} 
{U}_{\alpha}[I,T] \vert \psi \rangle 
\Vert dt < \infty
\label{eq:FT}
\eeq
where 
\beq
{V}_{\alpha} := {H} {\Phi}_{\alpha} -
{\Phi}_{\alpha} {H}_{\alpha} .
\label{eq:FU}
\eeq
The scattering operator for scattering from channel $\alpha$ to channel 
$\beta$ is the mapping from ${\cal H}_{\alpha} \to {\cal H}_{\beta}$
defined by 
\beq
S_{\alpha \beta} := {\Omega}^{\dagger}_{\alpha +}  
{\Omega}_{\beta -} .
\label{eq:FV}
\eeq
 
This can be compactly expressed in the two-Hilbert space formulation.  The 
asymptotic Hilbert space is the direct sum of all of the channel subspaces,
including all bound state channels:
\beq
{\cal H}_{\cal A} := \oplus_{\alpha \in {\cal A}} {\cal H}_{\alpha} .
\label{eq:FW}
\eeq
A two-Hilbert space injection operator ${\Phi}_{\cal A}$ is a mapping from 
${\cal H}_{\cal A}$ to ${\cal H}$ defined by
\beq
{\Phi}_{\cal A} := \sum_{\alpha \in {\cal A}} {\Phi}_{\alpha}
\label{eq:FX}
\eeq
where each ${\Phi}_{\alpha}$ acts on the subspace ${\cal H}_{\alpha}$
of ${\cal H}_{\cal A}$. 

There is a natural unitary representation of $ISL(2,C)$ on ${\cal H}_{\cal A}$
defined by 
\beq
{U}_{\cal A}[\Lambda ,Y] := \sum_{\alpha \in {\cal A}}
{U}_{\alpha}[\Lambda ,Y].   
\label{eq:FY}
\eeq
The two-Hilbert space wave operators can be expressed by the strong limits
\beq
\Omega_{\pm}({H}, {\Phi}_{\cal A}, {H}_{\cal A}):=
\lim_{t \to \pm \infty}  {U}^{\dagger}[I,T]{\Phi}_{\cal A} {U}_{\cal A}[I,T]
\label{eq:FZ}
\eeq
where $T=t\sigma_0$.

In what follows the dynamical model is assumed to have two 
Hilbert-space wave operators that exist and are asymptotically 
complete in the sense that the $S$ is unitary.  With these 
assumptions and some restrictions on the 
interactions \cite{wp} the two-Hilbert space 
wave operators are unitary operators from ${\cal H}_{\cal A} 
\to {\cal H}_{\cal A}$ satisfying the intertwining property:
\beq
{U}[\Lambda ,Y] 
\Omega_{\pm}({H}, {\Phi}_{\cal A}, {H}_{\cal A}) =
\Omega_{\pm}({H}, {\Phi}_{\cal A}, {H}_{\cal A})
{U}_{\cal A}[\Lambda ,Y]. 
\label{eq:FAA}
\eeq

\section{Scattering Equivalences}

For a fixed number of particles a scattering equivalence is
a unitary operator ${A}$ satisfying 
\beq
s-\lim_{t \to \pm \infty} ({A}-{I})U_0 [I,T] = 0
\label{eq:GA}
\eeq
for $T := t \sigma_0$.  The physical significance of the asymptotic condition
(\ref{eq:GA}) is that the unitary transformation ${A}$ transforms
the Hamiltonian in a manner that leaves the spectrum and scattering
observables unchanged, {\it without changing the representation of a free
particle} used to formulate the asymptotic condition.  

What this means is that if the relevant observables are $S$-matrix
elements and spectral properties, and the dynamics is defined by
adding interactions to a free-particle dynamics, there is large class
of interactions that give the same $S$-matrix elements and spectral
properties.

Specifically, if ${U}'[\Lambda ,Y] := {A}{U}[\Lambda ,Y]   
{A}^{\dagger}$,  condition (\ref{eq:GA}) is equivalent to 
\beq
\Omega_{\pm} ({H}', {\Phi}_{\cal A}', {H}_{\cal A} ) =
{A} \Omega_{\pm} ({H}, {\Phi}_{\cal A}, {H}_{\cal A} )
\label{eq:GB}
\eeq
with 
${\Phi}'_{\alpha_0} = {\Phi}_{\alpha_0}$
when $\alpha_0$ is the $N$-body breakup channel.

Eq. (\ref{eq:GB}) ensures that all $S$ matrix elements are preserved:
\[
S ({H}' , {\Phi}_{\cal A}' , {H}_{\cal A} ) =
\Omega^{\dagger}_{+} ({H}', {\Phi}_{\cal A}', {H}_{\cal A} )
\Omega_{-} ({H}', {\Phi}_{\cal A}', {H}_{\cal A} ) =
\]
\beq
\Omega^{\dagger}_{+} ({H}, {\Phi}_{\cal A}, {H}_{\cal A} )
\Omega_{-} ({H}, {\Phi}_{\cal A}, {H}_{\cal A} ) =
S ({H} , {\Phi}_{\cal A} , {H}_{\cal A} ).
\label{eq:GC}
\eeq
The distinction between minimally-charged bare particles and
``composite'' bare particles is relevant for generalizing the condition
(\ref{eq:GA}).  The special property of minimally-charged bare
particles is that they exist as asymptotically separated stable bare
particles.

The generalization of (\ref{eq:GA}) is easy to formulate.
First consider only those channels $\alpha_m$ where each asymptotic 
particle is a ``minimum-charge'' particle.   
Consider the class of unitary transformations $A$ on ${\cal H}$ 
with the property that they preserve the full $S$-matrix
\beq
{H}' = {A} {H} {A}^{\dagger}
\label{eq:GD}
\eeq
\beq
S = S' 
\label{eq:GE}
\eeq
where 
\beq
S :=
\Omega^{\dagger}_{+} ({H}, {\Phi}_{\cal A}, {H}_{{\cal A}} )
\Omega_{-} ({H}, {\Phi}_{\cal A}, {H}_{\cal{A}} ). 
\label{eq:GF}
\eeq
The equality of the scattering operators and asymptotic completeness 
imply 
\beq
\Omega_{+} ({H}, {\Phi}_{\cal A}, {H}_{\cal{A}} )
\Omega^{\dagger}_{+} ({H}', {\Phi}'_{}, {H}_{{\cal A}} )=
\Omega_{-} ({H}', {\Phi}_{\cal A}, {H}_{\cal{A}} )
\Omega^{\dagger}_{-} ({H}', {\Phi}'_{}, {H}_{{\cal A}} ).
\label{eq:GG}
\eeq
A sufficient condition for this to be true is 
\beq
\Omega_{\pm} ({H}', {\Phi}_{\cal A}, {H}_{\cal{A}} )=
A \Omega_{\pm} ({H}, {\Phi}_{}, {H}_{{\cal A}} )
\label{eq:GH}
\eeq
for {\it both} asymptotic conditions.  It is a non-trivial condition
that there is a single solution $A$ to (\ref{eq:GH}) for both
asymptotic conditions.   This is equivalent to the condition
\beq
s-\lim_{t \to \pm \infty} [A\Phi_{\cal A} - \Phi'_{\cal A}]
{U}_{\cal A}[I,t] = 0 .
\label{eq:GI}
\eeq
In general, given a unitary $A$, it is possible to define
$\Phi'_{\cal A} := A \Phi_{\cal A}$, however for minimal charge 
channels, $\alpha_m$, 
which have no asymptotic bound clusters, any reasonable 
model must also require 
\beq
\Phi_{\alpha_m} = \Phi'_{\alpha_m}.
\label{eq:GJ}
\eeq
This requirement puts a non-trivial condition on $A$ given by
\beq
s-\lim_{t \to \pm \infty} (A-I) \Phi_{\alpha_m}
{U}_{\alpha_m}[I,T] = 0 
\label{eq:GK}
\eeq
which must hold for each minimal-charge channel, $\alpha_m$, and 
both time limits.

For systems of a fixed number of particles, scattering equivalences are
used to relate tensor product representations of $ISL(2,C)$, which are
useful cluster limits of a satisfactory dynamical model, with
representations where the mass commutes with the spin and a maximal
set of functions of the non-interacting generators.  Both
representation are needed to combine interactions that appear in
different asymptotic configurations.  What makes this work is that the
$N$-particle Hilbert space can be factored into a tensor product of
subsystem spaces.

What is different for the models under consideration is that the tensor
product of subsystem Hilbert spaces is not the whole Hilbert space.
Specifically, for a partition $a$ of charges the Hilbert
space has the decomposition ${\cal H} = {\cal H}_a \oplus {\cal H}^a$ and 
Hilbert space for the tensor product of the subsystems defined  
by the partition $a$ is ${\cal H}_a$.  The construction in
\cite{wp} naturally leads to a scattering equivalence $A_a$
on ${\cal H}_a$.  In order to treat cluster properties 
${A}_a$ must be extended
to all of ${\cal H}$.  The obvious extension $\tilde{A}_{a}$ that preserves 
all of the required properties of ${A}_a$ is 
\beq
\tilde{A}_{a} := 
\left (
\begin{array}{cc} 
{A}_{a} & 0\\
0 & {I}^a 
\end{array}
\right )
\label{eq:GL}
\eeq
where ${A}_a: {\cal H}_a \to {\cal H}_a$ and ${I}^a$ is the 
identity on ${\cal H}^a$.   In addition, because ${\cal H}_{a \cap b}
\subseteq {\cal H}_a$, the ${\cal H}^a$ is not relevant for 
cluster properties.  In all that follows the symbol $A_a$ is used to 
denote both $A_a$ and the extension $\tilde{A}_{a}$.

When the Hilbert space is extended to include composite bare
particles, the range of ${\Phi}_{\alpha_m}$, corresponding to {\it
minimal charge channels}, is orthogonal to all of the subspaces
associated with composite bare particles.  The condition on the
scattering equivalences are that in addition to being unitary
operators that preserve all scattering matrix elements, they satisfy
the condition (\ref{eq:GK}) for each minimal-charge channel,
$\alpha_m$.

The $C^*$ algebra of asymptotic constants defined in \cite{wp} is
replaced by the a new $C^*$ algebra of operators $Z$ subject to the
asymptotic conditions:
\beq
s-\lim_{t \to \pm \infty} {Z} {\Phi}_{\alpha_m}{U}_{\alpha_m} [I,T] = 0
\label{eq:GN}
\eeq
\beq
s-\lim_{t \to \pm \infty} {Z}^{\dagger} {\Phi}_{\alpha_m}{U}_{\alpha_m} [I,T] = 0
\label{eq:GO}
\eeq
for each minimal-charge channel $\alpha_m$.
This algebra is completed by including the identity.  The importance
of this algebra is that unitary elements of this algebra are
scattering equivalences.  Operations on the algebra provide a
functional calculus for constructing new scattering equivalences
which are functions of non-commuting scattering equivalences. 

\section{Summary}
 
In the preceding sections the modifications of the fixed number of
particle construction \cite{wp} necessary to treat particle production were
discussed.  The first new feature is that the Hilbert space does not
factor into a tensor product of subsystems Hilbert spaces.  Instead, for
any decomposition into subsystems, there is a residual subspace ${\cal
H}^a$ of (\ref{eq:DA}) where the factorization is not compatible with
the bare particle content on these subspaces.  This led to
modifications of the formulation of cluster properties (\ref{eq:EB}),
the asymptotic dynamics (\ref{eq:EP},\ref{eq:FL}), and the structure
of the algebra of scattering equivalences
(\ref{eq:GK},\ref{eq:GL},\ref{eq:GN},
\ref{eq:GO}).  Charges were introduced to
replace particles and cluster properties were formulated with respect
to partitions of charges.  The partial ordering on charges had 
important consequences.  The most important was that subsequent 
refinements of clusters never affected the residual component of the 
Hilbert space. 
Subsequent refinements only acted non-trivially on the 
tensor product subspace associated with the preceding refinement.  
While in the general case the
addition of fictitious charges has no consequence, the requirement
that all particles have positive charge limits the class of theories
to theories with a bounded number of bare-particle degrees of freedom.
The theories consistent with this requirement have structures like the
Lee model and Isobar models.  The value of this restriction is the
existence of a meaningful few-body dynamics.  Specifically, the
dynamics of the $N$-charge system is determined by the dynamics of the 
$K<N$ charge systems up to $N$-charge interactions.
While the ultimate goal is to remove the restriction to positive charges, 
the bounded charge theories exhibit all of required properties.
 
The modifications discussed are adequate to allow the general methods
used in \cite{wp} to be extended to treat the class of models
discussed in this paper.  In the next three sections
the modifications to the general construction are illustrated with
an example in the three charge sector.

\section{The Two-Charge System} 

The general construction of a dynamical representation of the
Poincar\'e group satisfying cluster properties is inductive, starting
with the simplest system.  The construction is illustrated with a
theory having two minimally-charged bare particles and two additional
composite-bare particles.  In this example the induction starts with the 
two-charge sector.  The first step is to specify the bare particles of the
theory:

\begin{itemize} 

\item[a.] There are two minimally-charged bare particles, 
labeled $N$ and $\pi$, with charges $(q_1,q_2) =(1,0)$ and $(0,1)$ 
respectively.

\item[b.] There are two composite bare particles labeled 
$\Delta$ and $\rho$ with charges $(1,1)$ and $(0,2)$ respectively.
\end{itemize} 

The charge conservation condition means that in addition to 
reactions that preserve particle number, the following basic reactions  
that change particle number are also possible
\beq 
\pi + \pi \leftrightarrow \rho \qquad N + \pi \leftrightarrow \Delta .
\label{eq:HA}
\eeq 
The composite bare particles of this model do not necessarily 
correspond to stable physical particles.  That will be true only if there
are point eigenstates of the mass operator with the same charge as 
the bare particle.  In a given model there could be zero, one, or
several physical $\Delta$ or $\rho$ particles.  In addition, 
the dynamics could lead to new composite particles, such as
a composite system of two or more $N$ particles,  which have different 
charges than any composite bare particle.  These are analogous to
bound states in the $N$-particle case.
  
The second step is to choose a representation to label the states of
the bare particles of the theory.  In this example the single-particle
observables, $F_i$ are chosen as the three components of the linear momentum,
$\vec{p}$, and the helicity, $h=\hat{p} \cdot \vec{j}$.  Vectors in 
the bare particle Hilbert spaces for a
bare particle of mass $m$ and spin $j$ are represented by square
integrable functions, $\psi (\vec{p},\lambda)$, of the eigenvalues,
$\lambda$, of the helicity, ${h}:=\hat{p}
\cdot \vec{j}$ and $\vec{p}$ of the linear momentum:
\beq
\vert \psi \rangle :=  \sum_{\lambda=-j}^j \int
\vert \vec{p} ,\lambda; m , j \rangle\,  d^3 p\,  \psi (\vec{p},\lambda).
\label{eq:HB}
\eeq
The bare-particle Hilbert spaces are denoted by ${\cal H}_{N}$,
${\cal H}_{\pi}$, ${\cal H}_{\Delta}$, and ${\cal H}_{\rho}$.

There are three two-charge problems corresponding to the total charges
$(q_1,q_2)= (2,0),(1,1)$ and $(0,2)$.  The $(2,0)$ problem corresponds
to two $N$ particles, and is an ordinary two-body problem
which can be treated using the methods of \cite{wp}\cite{wp2}.  The model
Hilbert space for the $(1,1)$ and $(0,2)$ sectors each include a
composite (total charge $>1$) bare particle.

The Hilbert spaces for the $(1,1)$ and $(0,2)$ sectors are 
\beq
{\cal H}_{(1,1)} := 
({\cal H}_{m_N, j_N} \otimes {\cal H}_{m_\pi, j_\pi})
\oplus  {\cal H}_{m_{\Delta},j_{\Delta}}
\label{eq:HC}
\eeq
and
\beq
{\cal H}_{(0,2)} := 
({\cal H}_{m_{\pi_1}, j_{\pi_1}} \otimes {\cal H}_{m_{\pi_2}, j_{\pi_2}})
\oplus  {\cal H}_{m_{\rho},j_{\rho}}.
\label{eq:HD}
\eeq
The bare-particle spaces, ${\cal H}_{m,j}$, are irreducible
representation spaces \cite{bkwp}\cite{wp} of $ISL(2,C)$.  Non-interacting 
basis
vectors on the charge-two Hilbert spaces ${\cal H}$ have the general
form:
\beq
\left (
\begin{array}{c}
\vert \vec{p}_1,\lambda_1 ;m_1,j_1\rangle \otimes  \vert \vec{p}_2,\lambda_2 
;m_2, j_2 \rangle  \\
0 
\end{array}
\right )      
\label{eq:HE}
\eeq
in the two-particle sector, ${\cal H}_a$, and 
\beq
\left (
\begin{array}{c}
0 \\
\vert \vec{p}_1,\lambda_1 ;m_{c},j_{c} \rangle 
\end{array} 
\right )
\label{eq:HEA}
\eeq
in the one-particle sector, ${\cal H}^a$.

There is a non-interacting unitary representation ${U}_0[\Lambda ,Y]$
of $ISL(2,C)$ on ${\cal H}$ corresponding to (\ref{eq:CG}), defined by 
\beq
{U}_0 [\Lambda ,Y] :=
\left (
\begin{array}{cc}
{U}_1[\Lambda ,Y] \otimes {U}_2 [\Lambda ,Y] & 0 \\
0 & {U}_{c} [\Lambda ,Y]
\end{array}
\right ). 
\label{eq:HF}
\eeq
where 
\beq
{U}_i [\Lambda ,Y] \vert \vec{p} , \lambda; m_i,j_i \rangle =
\label{eq:HG}
\eeq
\beq
\sum_{\lambda'=-j}^j \vert \vec{p}\,', \lambda' ;m_i,j_i \rangle
\sqrt{\omega_{i} (p') \over \omega_{i} (p)} 
D^{j}_{\lambda' \lambda} [R_H (\Lambda ,p)]e^{i p' \cdot y} 
\label{eq:HH}
\eeq
with 
\beq
\omega_i(p) := \sqrt{m_i^2 + \vec{p} \cdot \vec{p}}
\label{eq:HI}
\eeq
\beq
p^{' \nu} := {1 \over 2} \mbox{Tr} ({\sigma}_{\nu}  \Lambda \sigma_{\mu}
\Lambda^{\dagger})p^{\mu} = \Lambda^{\nu}{}_{\mu} p^{\mu}   
\label{eq:HJ}
\eeq
\beq
R_H (\Lambda ,p):= H^{-1}(p_{\Lambda})\Lambda H(p)
\label{eq:HK}
\eeq
with $H(p)$ is the helicity boost \cite{wick} given by
\beq
H(p) := R(\hat{z} \to \hat{p}) B_c (\vert\, \vec{p}\, \vert \hat{z}). 
\label{eq:HL}
\eeq
The matrix $R(\hat{z} \to \hat{p})$ is the $SU(2)$ matrix 
corresponding to a rotation about the axis 
parallel to $\hat{\theta}={\hat{z}\times 
\hat{p} \over \vert \hat{z}\times \hat{p} \vert} $ through an angle $\theta$ 
define by $\cos (\theta) = \hat{z} \cdot \hat{p}$.  It rotates $\hat{z}$
into the direction parallel to $\vec{p}$.
This matrix has the form
\beq
R(\hat{z} \to \hat{p}) = \sigma_0 \cos ({\theta \over 2})
+ i {\theta} \cdot \vec{\sigma}  \sin ({\theta \over 2}).
\label{eq:HM}
\eeq
The matrix $B_c (\vert\, \vec{p}\, \vert \hat{z})$ is the $SL(2,C)$ 
matrix corresponding to a rotationless Lorentz transformation in the $\hat{z}$ 
direction given by 
\beq
B_c (\vert\, \vec{p}\, \vert \hat{z}) = \sigma_0 \cosh ({\eta \over 2})
+ {\sigma}_z  \sinh ({\eta \over 2})
\label{eq:HN}
\eeq
where the rapidity $\eta$ satisfies
\beq
\sinh (\eta) = {\vert \,\vec{p} \,\vert \over m_i} \qquad
\cosh (\eta) = {\omega_i (p) \over m_i}.
\label{eq:HO}
\eeq
The matrix $D^{j}_{\lambda' \lambda} [R]$ 
is the $SU(2)$ Wigner D-function of $R$.
The tensor product basis is not a useful basis for including interactions.
It is more useful to work in an equivalent basis which transforms 
irreducibly with respect to ${U}_0 [\Lambda ,Y]$.

Clebsch-Gordan coefficients \cite{wick}\cite{bkwp}\cite{wp} 
of $ISL(2,C)$ can be used to construct
linear combinations of the basis elements on the subspace ${\cal H}_1
\otimes {\cal H}_2$ that transform irreducibly with respect to
${U}_0 [\Lambda ,Y]$.  The form of the Clebsch-Gordan coefficients in the
helicity basis depends on the choice of degeneracy quantum numbers.
Wick \cite{wick} uses ``body-fixed'' single particle helicities
to label degeneracies.  In this model we use ``spin'' and ``orbital''
angular momentum labels that are more natural for formulating 
two-body interactions.  For this choice the Clebsch-Gordan 
coefficients are:
\[
\langle
\vec{p}_1,\lambda_1 ;m_{1},j_{1}: \vec{p}_2,\lambda_2 ;m_{2},j_{2} 
\vert
\vec{p}_{12},\lambda_{12} ;k_{12},j_{12}; l_{12},s_{12} \rangle =
\]
\[
\sum_{\lambda_1',\lambda_2',\lambda_{s12}',\mu_{l}'}
\delta (\vec{p}_{12} - \vec{p}_1 -\vec{p}_2 )
{\delta (k_{12} - k(\vec{p}_1,\vec{p}_2) ) \over k^2_{12} } \times
\]
\[
\vert {\partial (\vec{p}_{12}, \vec{k}_1 (\vec{p_1},\vec{p}_2))
\over \partial (\vec{p}_{1},\vec{p}_{2})}\vert^{1/2}   
Y^l_{\mu_l}(\hat{k}_{1})   
\times
\]
\[
D^{j_1}_{\lambda_1 ,\lambda_1'}[R_{HMW}(p_{12},k_1)]
D^{j_2}_{\lambda_2 ,\lambda_2'}[R_{HMW}(p_{12},k_2)] \times
\]
\beq
\langle j_1, \lambda_1', j_2, \lambda_2' \vert s_{12} ,\lambda_{s12'}
\rangle
\langle s_{12}, \lambda_{s12}', l, \mu_l' \vert j_{12} ,\lambda_{12}
\rangle
\label{eq:HP}
\eeq
where
\beq
R_{HMW}(p,k_i) :=H^{-1} (p_{i})H(p_{12}) B_c (k_i)
\label{eq:HQ}
\eeq
is a rotation obtained by composing a helicity-Melosh rotation with a
helicity-Wigner rotation \cite{bkwp}.  The other quantities in (\ref{eq:HQ})
are defined by:
\beq
{k}_{i}^{\nu} := {1 \over 2} \mbox{Tr} ({\sigma}_{\nu}  
H^{-1} (p_{12}) \sigma_{\mu}
(H^{-1}(p_{12}))^{\dagger} ) p_i^{\mu} 
\label{eq:HR}
\eeq
\beq
B_c (k_i) = \sigma_0 \cosh ({\eta_i \over 2})
+ \hat{k}_i \cdot \vec{\sigma}  \sinh ({\eta_i \over 2})
\label{eq:HS}
\eeq
and
\beq
\sinh (\eta_i) = {\vert \vec{k}_i \vert \over m_i} \qquad
\cosh (\eta_i) = {\omega_i (k_i) \over m_i} .
\label{eq:HT}
\eeq
In this coefficient the mass $m_{12}$ is related to the continuous variable 
$k_{12}^2$ by
\[
k_{12}^2 := 
\]
\beq
{{m}_{12}^4 +{m}_{1}^4 + 
m_2^4 - 2{m}_{1}^2 m_2^2 - 
2{m}_{12}^2 {m}_{1}^2 - 
2 {m}_{12}^2  m_2^2
\over 4 {m}_{12}^2} 
\label{eq:HU}
\eeq
which has a spectrum $\in [0,\infty]$.  The Jacobian is 
\beq
\vert {\partial (\vec{p}_{12}, \vec{k}_1(\vec{p_1}, \vec{p}_2\,))
\over \partial (\vec{p}_{1},\vec{p}_{2})}\vert=
{\omega_1 (k_1) \omega_2 (k_2)(\omega_1 (p_1)+ \omega_2 (p_2)) \over 
\omega_1 (p_1) \omega_p (p_2)(\omega_1 (k_1)+ \omega_2 (k_2))}.
\label{eq:HV}
\eeq

These Clebsch-Gordan coefficients define 
the irreducible non-interacting eigenstates
\beq
\vert
\vec{p}_{12},\lambda_{12} ;k_{12},j_{12}; l_{12},s_{12} \rangle 
\label{eq:HW}
\eeq
as linear superpositions of the tensor product states. 
The irreducible non-interacting eigenstates
transform as:
\[
U_1[\Lambda, Y] \otimes 
U_2[\Lambda, Y]\vert
\vec{p}_{12},\lambda_{12} ;k_{12},j_{12}; l_{12},s_{12} \rangle =
\]
\[
\sum_{\lambda_{12}'=-j_{12}}^{j_{12}} 
\vert \vec{p}_{12}\,',\lambda_{12}' 
;k_{12},j_{12}; l_{12},s_{12} \rangle \times
\]
\beq
\sqrt{\omega_{12} (p_{12}') \over \omega_{12} (p_{12})} 
D^{j_{12}}_{\lambda_{12}', \lambda_{12}} [R_H(\Lambda , p_{12})]
e^{i p' \cdot y}.
\label{eq:HX}
\eeq
This has the same structure as a single particle transformation,
except the non-interacting two-body invariant mass $m_{12}$ is 
replaced by the more convenient continuous
variable $k_{12}^2$.  The transformed four-momentum $p'$ is related to the
original momentum by (\ref{eq:HJ}). The quantum numbers $s_{12}$ and
$l_{12}$ are invariant degeneracy quantum numbers, which are needed
because multiple copies of the $m_{12}$, $j_{12}$ representation
appear in the tensor product.

The irreducible free eigenstates,
\beq
\left (
\begin{array}{c}
\vert
\vec{p}_{12},\lambda_{12} ;k_{12},j_{12}; l_{12},s_{12} \rangle  \\
0 
\end{array}
\right ) 
\label{eq:HY}
\eeq
and  
\beq
\left (
\begin{array}{c}
0 \\
\vert \vec{p}_c,\lambda_c ;m_c,j_c \rangle  
\end{array}
\right ),     
\label{eq:HZ}
\eeq
are a basis which can be used to solve the two-charge dynamics.

The dynamics is defined by adding an interaction, $V$,  to
the free mass operator, $M_{0}$  that commutes with and is independent 
of the $\vec{p}_{12}$ and $\lambda_{12}$. 
  
In the free-particle irreducible basis the non-interacting mass operator 
has the form
\beq
{M}_0 =
\left (
\begin{array}{cc} 
m_{12}  & 0 \\
0 & m_{c}  
\end{array} 
\right ) =
\left (
\begin{array}{cc} 
\sqrt{m_1^2 + k_{12}^2}+ \sqrt{m_{2}^2 + k_{12}^2} & 0 \\
0 & m_{c}  
\end{array} 
\right ) 
\label{eq:HAA}
\eeq
and the interaction is assumed to have a kernel of the form:
\beq
\langle \vec{p}, \lambda; \vec{j} \cdots \vert V \vert 
\vec{p}\,', \lambda'; \vec{j}' \cdots \rangle =
\label{eq:HAB}
\eeq
\beq
\delta (\vec{p} - \vec{p}\,') \delta_{j j'}
\delta_{\lambda \lambda'}
\left (
\begin{array}{cc}
\langle k_{12}, l_{12}, s_{12}  \Vert V^j  \Vert k_{12}', l_{12}', s_{12}'
\rangle & \langle  
k_{12}, l_{12}, s_{12} \Vert V^j \Vert m_c 
\rangle \\
\langle m_c  \Vert V^j \Vert k_{12}', l_{12}', s_{12}' \rangle &
\langle m_c \Vert V^j \Vert m_c \rangle
\end{array}
\right ) .
\label{eq:HAC}
\eeq 
The term 
$\langle m_c \Vert V^j \Vert m_c \rangle$ is a 
constant which could be absorbed in the bare mass, $m_c$. 

For an interaction of the form (\ref{eq:HAC}) the dynamical mass operator
\beq
{M}:= {M}_0 +{V} 
\label{eq:HAD}
\eeq
commutes with and is independent of $\vec{p}$ and ${h}$.  It
follows that ${M}, \vec{p}, {h}, {j}^2$ can be
simultaneously diagonalized by diagonalizing ${M}$ in the
free-particle irreducible basis (\ref{eq:HY},
\ref{eq:HZ}).  In this basis the mass 
eigenfunctions have the form
\beq
\langle \vec{p}, \lambda ; j , \cdots 
\vert \vec{p}\,' \lambda' ; m' ,j' \rangle  =
\delta (\vec{p} - \vec{p}\,') \delta_{j j'}
\delta_{\lambda \lambda'}
\left (
\begin{array}{c}
\langle k_{12} , l_{12} , s_{12}  \vert m', j \rangle \\
\langle m_c \vert m', j \rangle
\end{array}
\right )
\label{eq:HAE}
\eeq
where the components of the reduced  wave function, 
$\langle k_{12} ,l_{12},s_{12}  \vert 
m, j \rangle$ and 
$\langle m_c \vert m, j \rangle$, are solutions of the coupled
equations:
\[
(m - M_0 )\langle k_{12},l_{12},s_{12} \vert m, j \rangle =
\]
\[
\sum_{l'_{12},s'_{12}} \int_0^\infty  \langle k_{12},l_{12},s_{12} 
\Vert V^j \Vert k_{12}',l_{12}',s_{12}' \rangle k_{12}^{\prime 2} dk_{12}' ,
\langle k_{12}',l_{12}',s_{12}' \vert m , j \rangle
+ 
\]
\beq
\langle k_{12},l_{12},s_{12} \Vert V^j \Vert m_c \rangle 
\langle m_c \vert m, j \rangle,
\label{eq:HAF}
\eeq
\[
(m - m_c )\langle m_c \vert m, j \rangle =
\]
\[
\sum_{l_{12}',s_{12}'} \int_0^\infty  
\langle m_c  \Vert V^j \Vert k_{12}',l_{12}',s_{12}' \rangle 
k_{12}^{\prime 2} dk_{12}'
\langle k_{12}',l_{12}',s_{12}' \vert m, j \rangle + 
\]
\beq
\langle m_c \Vert V^j \Vert m_c \rangle
\langle m_c \vert m, j \rangle
\label{eq:HAG}
\eeq
for the eigenvalue $m$.  These equations can be
combined into a single dynamical equation for 
$\langle k_{12}',l_{12}',s_{12}' \vert
m, j \rangle$: 
\[
(m -m_0 )\langle k_{12},l_{12},s_{12} \vert m, j \rangle =
\]
\beq
\int_0^{\infty}\sum_{s'_{12},l'_{12}} 
\langle k_{12},l_{12},s_{12} 
\vert K^j  \vert k_{12}',l_{12}',s_{12}' \rangle  k_{12}^{\prime 2} dk_{12}'  
\langle k_{12}',l_{12}',s_{12}' \vert m, j \rangle
\label{eq:HAH}
\eeq
where 
\[
\langle k_{12},l_{12},s_{12} 
\vert K^j  \vert k_{12}',l_{12}',s_{12}' \rangle := 
\]
\beq
\langle k_{12},l_{12},s_{12} 
\Vert V^j \Vert k_{12}',l_{12}',s_{12}' \rangle 
+ {\langle k_{12},l_{12},s_{12} \Vert V^j \Vert m_c \rangle 
\langle m_c  \Vert V^j \Vert q_{12}',l_{12}',s_{12}' \rangle
\over m -m_c - \langle m_c \Vert V^j \Vert m_c \rangle}.
\label{eq:HAI}
\eeq
The component $\langle m_c \vert m, j \rangle$ can be obtained by 
quadrature:
\[
\langle m_c \vert m, j \rangle = 
\]
\beq
\int_0^{\infty}\sum_{s'_{12},l'_{12}}  
{ \langle m_c  \Vert V^j \Vert k_{12}',l_{12}',s_{12}' \rangle
\over m -m_c - \langle m_c \Vert V^j \Vert m_c \rangle}
k_{12}^{\prime 2} dk_{12}'
\langle k_{12}',l_{12}',s_{12}' \vert m, j \rangle .
\label{eq:HAJ}
\eeq
For scattering states equations (\ref{eq:HAH},\ref{eq:HAJ}) must 
be solved with incoming or outgoing 
asymptotic conditions, $m \to m \pm i 0+$.
This dynamical equation is of comparable difficulty to solving 
the two-body Lippmann-Schwinger equation. 

For a self-adjoint ${M}$ with a well-behaved short-ranged
interaction $V$  the simultaneous eigenstates (\ref{eq:HAE}),
$\vert \vec{p}, \lambda ; m
,j \rangle$ of ${M}$, $\vec{p}=\vec{p}_0$, ${h}={h}_0$,
${j}^2 = {j}_0^2$ are a complete set of eigenstates that
transform irreducibly with respect to a dynamical representation
${U}[\Lambda ,Y]$ of $ISL(2,C)$.  The transformation properties of
these eigenstates follow because the operators ${M}_0$,
$\vec{p}$, $\vec{j}$,$i
\vec{\nabla}_p$ and ${M}$, $\vec{p}$, $\vec{j}$,$i \vec{\nabla}_p$
have identical commutation relations. Note that in both 
cases the partial $p$-derivatives 
are performed holding the helicity constant.

It follows that
${U}[\Lambda ,Y]$ is defined in the basis (\ref{eq:HAE}) by
\beq
{U}[\Lambda ,Y] \vert \vec{p}, \lambda ; m ,j \rangle = 
\sum_{\lambda'=-j}^j \vert \vec{p}\,', \lambda' ;m,j \rangle
\sqrt{\omega_{m} (p') \over \omega_{m} (p)} 
D^{j}_{\lambda' \lambda} [R_H (\Lambda ,p)]e^{i p' \cdot y} 
\label{eq:HAK}
\eeq
where 
\beq
p^{' \nu} := {1 \over 2} \mbox{Tr} ({\sigma}_{\nu}  \Lambda \sigma_{\mu}
\Lambda^{\dagger})p^{\mu} = \Lambda^{\nu}{}_{\mu} p^{\mu}   
\label{eq:HAL}
\eeq
with $p^0:= \sqrt{\vec{p}\cdot \vec{p} +m^2}$.

This is identical to the transformation properties of a single particle,
except the {\it mass parameter is the eigenvalue $m$ of the mass 
operator (\ref{eq:HAD})}.

The action of ${U}[\Lambda ,Y]$ on an arbitrary state
with wave-function $\psi (\vec{p} , \lambda)$  is given by 
completeness:
\[
{U}[\Lambda ,Y] \vert \psi \rangle = 
\]
\beq
\sum_{\lambda,\lambda'=-j}^j \int \vert \vec{p}\,', \lambda' ;m,j \rangle
\sqrt{\omega_{m} (p') \over \omega_{m} (p)} 
D^{j}_{\lambda' \lambda} [R_H (\Lambda ,p)]e^{i p' \cdot y} 
\psi_{m,j}(\vec{p} ,\lambda) d^3 p .
\label{eq:HAM}
\eeq

Solutions of the dynamical equations with $m < m_1 + m_2$ in the
pure point spectrum of ${M}$ correspond to stable $\rho$ or $\Delta$ 
particles.
Solutions with $m \geq m_1 + m_2$ in the absolutely continuous
spectrum of ${M}$ correspond to scattering eigenstates.  We assume
that all of the eigenstates fall into one of these two
classifications.  In addition, we assume that the incoming and
outgoing wave scattering solutions each span the subspace of
${\cal H}$ orthogonal to the bound state subspace.

This representation satisfies algebraic cluster properties
because the
operator $U[\Lambda ,Y]$ is a function of single bare-particle operators 
and $V$. In the limit that $V$ vanishes this representation becomes
the non-interacting representation ${U}_0[\Lambda ,Y]$.

This completes the $\vert {\cal N}\vert=2$ construction for the 
charge sector $(1,1)$ or $(0,2)$.

\section{The $\{ 2,1\}$ sector}

The next step in the construction of the dynamics in the three-charge
sector is to consider the problem of two interacting charges and a
spectator charge.  This problem defines the asymptotic behavior of the
three-charge system in the limit that one charge is asymptotically
separated from an interacting pair. This form is used in the 
mathematical formulation of cluster properties.

To be specific consider the three-charge system with 
$(q_1,q_2) = (1,2)$.  For this set of 
charge quantum numbers the Hilbert space is  
\beq
{\cal H} = ( {\cal H}_{N} \otimes {\cal H}_{\pi_1} \otimes {\cal H}_{\pi_2}) 
\oplus ({\cal H}_{\Delta_1} \otimes {\cal H}_{\pi_2})
\oplus ({\cal H}_{\Delta_2} \otimes {\cal H}_{\pi_1})
\oplus ({\cal H}_{N} \otimes {\cal H}_{\rho}).
\eeq

The non-trivial partitions ${\cal P}$ of the minimal charges of this
system are $a=(N)(\pi_1)(\pi_2)$,
$(N,\pi_1)(\pi_2)$,$(N,\pi_2)(\pi_1)$, and $(\pi_1,\pi_2)(N)$.  For
each partition $a\in {\cal P}$ there is a decomposition of the 
Hilbert space of the form (\ref{eq:DA}).
For example, for $a= (N)(\pi_1)(\pi_2) $ the orthogonal subspaces are
\beq
{\cal H}_a =  {\cal H}_{N} \otimes {\cal H}_{\pi_1} \otimes {\cal H}_{\pi_2} 
\label{eq:IA}
\eeq
\beq
{\cal H}^a =
({\cal H}_{\Delta_1} \otimes {\cal H}_{\pi_2})
\oplus ({\cal H}_{\Delta_2} \otimes {\cal H}_{\pi_1})
\oplus ({\cal H}_{N} \otimes {\cal H}_{\rho})
\label{eq:IB}
\eeq
and for $a= (N, \pi_1)(\pi_2)$ the orthogonal subspaces are
\[
{\cal H}_a =  ({\cal H}_{N} \otimes {\cal H}_{\pi_1} 
\otimes {\cal H}_{\pi_2})  
\oplus ({\cal H}_{\Delta_1} \otimes {\cal H}_{\pi_2})
\]
\[
=[({\cal H}_{N} \otimes {\cal H}_{\pi_1}) 
\oplus {\cal H}_{\Delta_1}] \otimes {\cal H}_{\pi_2}
\]
\beq
={\cal H}_{(N\pi_1)} \otimes {\cal H}_{\pi_2}
\label{eq:IC}
\eeq
\beq
{\cal H}^a = ({\cal H}_{\Delta_2} \otimes {\cal H}_{\pi_1})
\oplus ({\cal H}_{N} \otimes {\cal H}_{\rho})
\label{eq:ID}
\eeq
with similar expressions for the two other non-trivial partitions $a$.

The goal of the three-body construction is to find a ${U}[\Lambda
,Y]$ that asymptotically factorizes into a tensor product of subsystem
representations when charges in different clusters of $a$ are
separated.  As mentioned in section 5, the 
formulation of cluster properties differs from
fixed-particle number case because ${U}[\Lambda,Y]$ acts on ${\cal
H}$ while $\otimes_i {U}_{a_i}[\Lambda,Y] $ acts on ${\cal H}_a$. 

Following (\ref{eq:EP}) it is useful to extend the asymptotic forms 
on ${\cal H}_a$ to operators on ${\cal H}$: 
\[
{U}_{(N)(\pi_1)(\pi_2)}[\Lambda ,Y] :=  
\]
\beq
\left (
\begin{array}{cc}
{U}_{N}[\Lambda ,Y]\otimes
{U}_{\pi_1}[\Lambda ,Y]\otimes
{U}_{\pi_2}[\Lambda ,Y] & 0 \\
0 & I
\end{array}
\right ) 
\label{eq:IE}
\eeq
for $a=(N)(\pi_1)(\pi_2)$ and 
\[
{U}_{(N\pi_1 )(\pi_2)}[\Lambda ,Y] :=  
\]
\beq
\left (
\begin{array}{cc}
{U}_{(N\pi_1)}[\Lambda ,Y]\otimes
{U}_{\pi_2}[\Lambda ,Y] & 0 \\
0 & I 
\end{array}
\right ) 
\label{eq:IF}
\eeq
for $a=(N\pi_1) (\pi_2)$, with similar expression for the two-cluster
partitions $(N\pi_2)(\pi_1)$ and $(N)(\pi_1 \pi_2)$.  These operators are
tensor products of subsystem representations of $ISL(2,C)$ on ${\cal H}_a$
and are extended so they are the identity on the subspace ${\cal H}^a$. 

In the first expression the factors ${U}_{\pi_i}[\Lambda ,Y]$ and
${U}_{N}[\Lambda ,Y]$ are single particle irreducible
representations. The factor ${U}_{(N\pi_1)}[\Lambda ,Y]$ in the
second expression is the {\it interacting} two-charge representation
constructed (\ref{eq:HAM}) in the previous section.  It acts on the
two-charge space ${\cal H}_{(N\pi_1)} := ({\cal H}_{N} \otimes {\cal
H}_{\pi_1})
\oplus {\cal H}_{\Delta_1}$.  

To formulate the cluster condition (\ref{eq:EB}), 
for each $a\in {\cal P}$, define
${\Pi}_a$ to be the orthogonal projector onto the subspace ${\cal
H}_a$ of ${\cal H}$:
\beq
{\Pi}_a :=
\left (
\begin{array}{cc}
{I} & 0 \\
0 & 0 
\end{array}
\right ) 
\label{eq:IG}
\eeq
and the cluster translation operators
\beq
{T}_a (Y_1, \cdots, Y_{n_a} )
:=
\left (
\begin{array}{cc}
{U}_{a_1}[I,Y_1] \otimes \cdots \otimes {U}_{a_{n_a}}[I,Y_{n_a}]  
& 0 \\
0 & 0 
\end{array}
\right )  
\label{eq:IH}
\eeq
which independently translate the charges in each cluster of the
partition $a$ by displacements $y_1 \cdots y_{n_a}$ where $(Y_i :=
y_i^{\mu}\sigma_{\mu})$.  In this example $a=(N\pi_1)(\pi_2)$,
$(N\pi_2)(\pi_1)$,$(\pi_1\pi_2)(N)$, or $(N)(\pi_1)(\pi_2)$.
Formally, for the partition $a=(N,\pi_1)(\pi_2)$, the formulation of
cluster properties (\ref{eq:EB}) for the three-charge problem is:
\beq
\lim_{(y_1-y_2)^2 \to \infty}
\Vert [{U}[\Lambda ,Y] - {U}_{(N\pi_1)(\pi_2)} [\Lambda ,Y] ]
{T}_{(N\pi_1)(\pi_2)}(Y_{1}, 
Y_{2} ){\Pi}_{(N\pi_1)(\pi_2)}  \vert \psi \rangle \Vert = 0
\label{eq:II}
\eeq
where the limit corresponds to a large space-like separation.  

The projection operator ${\Pi}_{(N\pi_1)(\pi_2)}$ projects on the 
subspace of the Hilbert space where independent translations of the 
$(N\pi_1)$ and $\pi_2$ subsystems are defined.

One term that is not eliminated by the projection operator is the part
of ${U}[\Lambda ,Y]$ that maps ${\cal H}_{(N\pi_1)(\pi_2)}$ to ${\cal
H}^{(N\pi_1)(\pi_2)}$.  This contribution involves the interactions of
the general type $\rho \leftrightarrow \pi_1 + \pi_2$ and $\Delta_2
\leftrightarrow N + \pi_2$.  These should vanish in the limit that
$\pi_2$ is asymptotically separated from the $N\pi_1$ system 
provided the interactions have sufficiently short range.
Condition (\ref{eq:II}) provides a range condition on these production
interactions.

In the three and many-charge system it is also necessary to consider
cluster properties for sequential limits.  In the fixed charge case
sequential limits require care because different cluster operators are
defined on different subspaces.  To illustrate the problem consider
the following three limits:

\begin{itemize} 
\item [1.] First take the limit corresponding to the partition of 
charges $(N\pi_2)(\pi_1)$, followed by $(N \pi_1)(\pi_2)$.

\item [2.] First take the limit corresponding to the partition of 
charges $(N\pi_1)(\pi_2)$, followed by $(N \pi_2)(\pi_1)$.

\item [3.] Take the limit corresponding to the partition of 
charges $(N)(\pi_1)(\pi_1)$.
\end{itemize}

Intuitively one expects that all three limits should give the free
dynamics on ${\cal H}_{(N)(\pi_1)(\pi_2)}$.  The problem is that 
the first two limits are
defined on the larger subspaces ${\cal H}_{(N \pi_1)(\pi_2)}
\supset {\cal H}_{(N)(\pi_1)(\pi_2)}$ and
${\cal H}_{(N\pi_2)(\pi_1)} \supset {\cal H}_{(N)(\pi_1)(\pi_2)}$ 
respectively.  

The projection and translation operators that appear in the 
asymptotic condition in each of these three cases are: 
\beq
{T}_{(N\pi_2)(\pi_1)}(Y_{N\pi_2},Y_{\pi_1}){\Pi}_{(N\pi_2)(\pi_1)}
{T}_{(N\pi_1)(\pi_2)}(Y_{N\pi_1},Y_{\pi_2}){\Pi}_{(N\pi_1)(\pi_2)},
\label{eq:IJ}
\eeq
\beq
{T}_{(N\pi_1)(\pi_2)}(Y_{N\pi_1},Y_{\pi_2}){\Pi}_{(N\pi_1)(\pi_2)}
{T}_{(N\pi_2)(\pi_1)}(Y_{N\pi_2},Y_{\pi_1}){\Pi}_{(N\pi_2)(\pi_1)},
\label{eq:IK}
\eeq
and
\beq
{T}_{(N)(\pi_1)(\pi_2)}(Y_{N},Y_{\pi_1},Y_{\pi_2})
{\Pi}_{(N)(\pi_1)(\pi_2)}
\label{eq:IL}
\eeq
respectively.  

In the representation used in this example, if all of the cluster 
displacements, $Y_i$, are space vectors with no time component, then 
the translation operators have no two-body terms and
\[
{T}_{(N\pi_1)(\pi_2)}(Y_{N\pi_1},Y_{\pi_2})
{T}_{(N\pi_2)(\pi_1)}(Y'_{N\pi_2},Y'_{\pi_1})
=
\]
\beq
{T}_{(N)(\pi_1)(\pi_2)} (Y_{N\pi_1}+Y_{N\pi_2}',
Y_{N\pi_1}+Y_{\pi_1}',Y_{\pi_2}+Y_{N\pi_2}')
\label{eq:IM}
\eeq
and
\beq
{\Pi}_{(N\pi_1)(\pi_2)}{\Pi}_{(N\pi_2)(\pi_1)}
={\Pi}_{(N)(\pi_1)(\pi_2)}
\label{eq:IN}
\eeq
etc. In this case all three limits have the same form.  For the case
that the relative displacements are space-like, but the individual
displacements have non-vanishing time components, the space of initial
vectors differs in all three cases.  For example,
${T}_{(N\pi_1)(\pi_2)}$ has components that map vectors from
${\cal H}_{\Delta_1} \otimes {\cal H}_{\pi_2}$ to ${\cal
H}_{(N)(\pi_1)(\pi_2)}$, while the for the other two configurations
the range of the initial projections are orthogonal to the subspace
${\cal H}_{\Delta_1} \otimes {\cal H}_{\pi_2}$.

In this case the contribution of the subspace ${\cal H}_{\Delta_1}
\otimes {\cal H}_{\pi_2}$ can be eliminated at the outset by insisting
that the sequential limits should give the same result only on the
common subspace where all of the limits are defined.  Effectively this
means that the limit should only be applied to vectors in the range 
of ${\Pi}_{(N)(\pi_1)(\pi_2)}$.

Mathematically this means this equation (\ref{eq:IJ}) should be replaced by:
\[
{T}_{(N\pi_2)(\pi_1)}(Y_{N\pi_2},Y_{\pi_1}){\Pi}_{(N\pi_2)(\pi_1)}
{T}_{(N\pi_1)(\pi_2)}(Y_{N\pi_1},Y_{\pi_2}){\Pi}_{(N\pi_1)(\pi_2)}\to
\]
\beq
{T}_{(N\pi_2)(\pi_1)}(Y_{N\pi_2},Y_{\pi_1}){\Pi}_{(N)(\pi_1)(\pi_2)}
{T}_{(N\pi_1)(\pi_2)}(Y_{N\pi_1},Y_{\pi_2}){\Pi}_{(N)(\pi_1)(\pi_2)}.
\label{eq:IO}
\eeq

Cluster properties in the two-charge sector allow the replacement of  
\beq
{T}_{(N\pi_2)(\pi_1)}(Y_{N\pi_2},Y_{\pi_1}){\Pi}_{(N)(\pi_1)(\pi_2)}
{T}_{(N\pi_1)(\pi_2)}(Y_{N\pi_1},Y_{\pi_2})
{\Pi}_{(N)(\pi_1)(\pi_2)} 
\label{eq:IP}
\eeq
by
\[
{T}_{(N)}(Y_{N\pi_2}+ Y_{N\pi_1})
{T}_{(\pi_1)}(Y_{\pi_1}) 
{T}_{(\pi_2)}(Y_{N\pi_2} + Y_{\pi_2})
\times 
\]
\beq
{\Pi}_{(N)(\pi_1)(\pi_2)}
{T}_{(N\pi_1)}(Y_{N\pi_1})
{\Pi}_{(N)(\pi_1)(\pi_2)} 
\label{eq:IQ}
\eeq
which is equivalent to formulating the cluster limit with the 
non-interacting translation operators.   Similar remarks apply for 
representations where the dynamical translation operators have 
interactions.

This shows that by projecting on the largest subspace where all
cluster translations are defined, cluster properties can be formulated
in a manner that is similar to the fixed number of particle case.
These observations generalize to sequential limits of systems with 
more than three charges.

The tensor product representations constructed satisfy cluster 
properties by definition.   These tensor product representations 
are important building blocks for the full three-charge dynamics. 

For each of the $2+1$ charge problems discussed, there is an alternate 
construction that leads to the same $S$-matrix.  

To see this recall that the irreducible free-particle basis constructed using 
the Clebsch-Gordan coefficients in (\ref{eq:HY}) have the form:
\beq
\left ( 
\begin{array}{c}
\vert p_{12} \lambda_{12} ; k_{12} ,j_{12}, l_{12},s_{12} \rangle \\
0
\end{array} 
\right ) 
\quad
\left ( 
\begin{array}{c}
0 \\
\vert p_{12}, \lambda_{12} ; ,m_c ,j_{c}  \rangle 
\end{array} 
\right ). 
\label{eq:IR}
\eeq
The tensor product of this basis with a spectator basis defines 
a basis on the subspace ${\cal H}_{(12)(3)}$ of ${\cal H}$:
\beq
\left ( 
\begin{array}{c}
\vert p_{12} \lambda_{12} ; k_{12} ,j_{12}, l_{12},s_{12} \rangle
\vert p_3 \lambda_3 ; m_3, j_3 \rangle \\
0
\end{array} 
\right ) 
\label{eq:IS}
\eeq
\beq
\left ( 
\begin{array}{c}
0 \\
\vert p_{12}, \lambda_{12} ; ,m_c ,j_{c}  \rangle 
\vert p_3 \lambda_3 ; m_3, j_3 \rangle
\end{array} 
\right ) .
\label{eq:IT}
\eeq
Since (\ref{eq:IS},\ref{eq:IT}) 
are tensor products of irreducible representations, 
the Clebsch-Gordan coefficients can be used to 
transform the tensor product basis to a three charge irreducible basis:
\beq
\left ( 
\begin{array}{c}
\vert p \lambda ; q_{(12)(3)}, j_{(12)(3)} , L_{(12)(3)},S_{(12)(3)},
k_{12} ,j_{12}, l_{12},s_{12},m_3,j_3 \rangle \\
0
\end{array} 
\right ) 
\label{eq:IU}
\eeq
\beq
\left ( 
\begin{array}{c}
0 \\
\vert p, \lambda ;q_{(12)(3)}, j_{(12)(3)} , L_{(12)(3)},S_{(12)(3)}
,m_c ,j_{c},m_3,j_3 \rangle 
\end{array} 
\right ) 
\label{eq:IV}
\eeq

The free two-charge mass operator is a multiplication 
operator in each of the representations (\ref{eq:IS}, \ref{eq:IT}) 
and (\ref{eq:IU},\ref{eq:IV}): 
\beq
{M}_{0:12} =
\left (
\begin{array}{cc} 
\sqrt{m_1^2 + k_{12}^2}+ \sqrt{m_{2}^2 + k_{12}^2} & 0 \\
0 & m_{c}  
\end{array} 
\right ). 
\label{eq:IW}
\eeq
Natural extensions of the 
two-charge interactions (\ref{eq:HAC}) are defined 
in each of the representation (\ref{eq:IS},\ref{eq:IT}) and
(\ref{eq:IU},\ref{eq:IV}) by:
\[
\langle \vec{p}_{12}, \lambda_{12}; \vec{j}_{12} 
\cdots \vec{p}_3, \lambda_3, m_3, j_3    \vert v \vert 
\vec{p}_{12}\,', \lambda_{12}'; \vec{j}_{12}' 
\cdots \vec{p}_3\,', \lambda_3', m_3 ',j_3' 
\rangle =
\]
\beq
\delta (\vec{p}_{12} - \vec{p}_{12}\,') \delta_{\lambda_{12} \lambda'_{12}}
\delta_{j_{(12)},j_{(12)}'}
\delta (\vec{p}_{3} - \vec{p}_{3}\,') \delta_{\lambda_{3} \lambda'_{3}}
\delta_{j_{3},j_{3}'}
\times
\]
\[
\left (
\begin{array}{cc}
\langle k_{12}, l_{12}, s_{12}  \Vert V^{j_{12}}  \Vert k_{12}', l_{12}', s_{12}'
\rangle & \langle  
k_{12}, l_{12}, s_{12} \Vert V^{j_{12}} \Vert m_c 
\rangle \\
\langle m_c  \Vert V^{j_{12}} \Vert k_{12}', l_{12}', s_{12}' \rangle &
\langle m_c \Vert V^{j_{12}} \Vert m_c \rangle
\end{array}
\right ) .
\label{eq:IZ}
\eeq
and 
\[
\langle \vec{p}, \lambda; \vec{j} \cdots \vert \bar{v} \vert 
\vec{p}\,', \lambda'; \vec{j}' \cdots \rangle =
\]
\beq
\delta (\vec{p} - \vec{p}\,') \delta_{\lambda \lambda'}
\delta_{j_{(12)(3)},j_{(12)(3)}'}
{\delta( q_{(12)(3)}-q_{(12)(3)}')\over q_{(12)(3)}^2} \times
\label{eq:IX}
\eeq
\beq
\delta_{L_{(12)(3)},L_{(12)(3)}'}
\delta_{S_{(12)(3)},S_{(12)(3)}'}
\delta_{j_{(12)},j_{(12)}'}
\times
\]
\[
\left (
\begin{array}{cc}
\langle k_{12}, l_{12}, s_{12}  \Vert V^{j_{12}}  \Vert k_{12}', l_{12}', s_{12}'
\rangle & \langle  
k_{12}, l_{12}, s_{12} \Vert V^{j_{12}} \Vert m_c 
\rangle \\
\langle m_c  \Vert V^{j_{12}} \Vert k_{12}', l_{12}', s_{12}' \rangle &
\langle m_c \Vert V^{j_{12}} \Vert m_c \rangle
\end{array}
\right ) .
\label{eq:IY}
\eeq 
 
Since both of the above interactions only differ in the choice 
of multiplicative delta functions, which are asymptotically 
equivalent, it follows that both
\beq
M_{12}:= M_{0:12} +v \qquad \bar{M}_{12} := M_{0:12} + \bar{v}
\label{eq:IAA}
\eeq
give the same two-charge mass eigenvalues and S-matrix elements. 
These are {\it different} operators, because $M_{12}$ 
commutes with $\vec{p}_3$ and 
$\bar{M}_{12}$ commutes with $\vec{q}_{(12)(3)}$, which is 
defined analogously to (\ref{eq:HR}).
The operator 
$\bar{M}_{12}$ has
the additional important property that it commutes with $\vec{p}$, 
$j_{(12)(3)}$, and $h_{(12)(3)}$ and is independent of $\vec{p}$ and 
$h_{(12)(3)}$.

The interesting property is that the interaction $\bar{v}$ satisfies
$[\bar{v},p_3]\not=0$.
This means that if $\bar{v}$ is a short ranged interaction, then 
it {\it vanishes} in the limit that the {\it spectator} 
particle 3 is translated infinitely 
far from the interacting 12 pair.  This is {\it not} the expected behavior 
when a non-interacting spectator is asymptotically separated from 
an interacting pair of particles.   This type of violation of cluster 
properties is characteristic of how cluster properties typically fail
in improperly formulated relativistic models. 
 
The dynamics given by $M_{12}$ and $\bar{M}_{12}$ are both defined on
the subspace ${\cal H}_a$ of ${\cal H}$, give the same spectral
properties and $S$-matrix and are thus related by a unitary scattering
equivalence $A_a$ on ${\cal H}_a$.  Since the interaction $v$ in the
tensor product representation satisfies cluster properties by
construction, it follows that scattering equivalences do not
necessarily preserve cluster properties.
The simplest way to construct the unitary 
transformation $A_a$ is to use the Clebsh-Gordan coefficients
of $ISL(2,C)$  
to transform the eigenstates of $M_{12}$, which transform like a
tensor product of an interacting two-charge representation and 
a spectator representation, to a superposition of irreducible 
representations of $ISL(2,C)$.  The eigenstates 
\beq
\vert p_{12}, \lambda_{12},m_{12}, j_{12}; \vec{p}, \lambda_3 ;m_2 ,j_3
\rangle,
\label{eq:IAB}
\eeq
which transform as a product of irreducible representations with respect
to ${U}_{(12)}[\Lambda ,Y]\otimes  {U}_{(3)}[\Lambda ,Y]$, 
are mapped to irreducible eigenstates of the form
\beq
\vert \vec{p}, \lambda; \tilde{q}_{(12)(3)},\vec{j},
\tilde{L}_{(12)(3)}, \tilde{S}_{(12)(3)}, m_{12}, j_{12}, \rangle  
\eeq
where $\tilde{q}^2_{(12)(3)}$ is related to the mass 
operator, $M_{(12)(3)}$,  of this irreducible representation by  
\[
\tilde{q}^2_{(12)(3)}=
\]
\beq
{{M}_{(12)(3)}^4 +{M_{12}}^4 + 
m_3^4 - 2{M_{12}}^2 m_3^2 - 
2{M}_{(12)(3)}^2 {M_{12}}^2 - 
2 {M}_{(12)(3)}^2  m_3^2
\over 4 {M}_{(12)(3)}^2} 
\label{eq:IAC}
\eeq
where $M_{12}$ is the {\it interacting} two-body mass operator.
This irreducible representation is obtained by first solving the 
two-charge problem followed by using Poincar\'e Clebsch-Gordan 
coefficients to construct a superposition of irreducible representations.

The mass operator $M_{(12)(3)}$ has the form 
\beq
M_{(12)(3)}= 
\sqrt{M_{12}^2 + \tilde{q}_{(12)(3)}^2 } +
\sqrt{m_{3}^2 + \tilde{q}_{(12)(3)}^2 }, 
\label{eq:IADD}
\eeq
which is the invariant mass operator associated with the tensor product 
$U_{(12)}[\Lambda ,Y]\otimes U_3[\Lambda ,Y]$ of the interacting 
two-body representation and the spectator representation
on ${\cal H}_{(12)(3)}$.  

It is also possible to use the operator $\bar{M}_{12}$ to construct
a three-charge mass operator in the barred representation:
\beq
\bar{M}_{(12)(3)}= 
\sqrt{\bar{M}_{12}^2 + q_{(12)(3)}^2 } +
\sqrt{m_{3}^2 + q_{(12)(3)}^2 } 
\label{eq:IAD}
\eeq
where in this case $q_{(12)(3)}^2$ is the non-interacting operator
that replaces the mass in the Clebsch-Gordan coefficients for the
non-interacting three charge system, (\ref{eq:IU}).  The operators
$\bar{M}_{(12)(3)}$, $j^2_{(12)(3)}=j^2_0$, $\vec{p}_{(12)(3)}=
\vec{p}_0$, $h_{(12)(3)}=h_0$, satisfy the same commutation relations as  
${M}_{0(12)(3)}$, $j^0_{(12)(3)}=j^0_0$, $\vec{p}_{(12)(3)}=
\vec{p}_0$, $h_{(12)(3)}=h_0$ where ${M}_{0(12)(3)}$ is the non-interacting 
invariant mass of the three charge system.  It follows that simultaneous 
eigenstates of both sets of operators have the same Poincar\'e transformation
properties if the eigenvalues of ${M}_{0(12)(3)}$ are 
replaced by the eigenvalues of $\bar{M}_{0(12)(3)}$.  This lead to 
a complete set of eigenstates
\beq
\overline{\vert \vec{p}, \lambda; {q}_{(12)(3)},\vec{j},
\tilde{L}_{(12)(3)}, \tilde{S}_{(12)(3)}, \bar{m}, j_{12}} \rangle  
\label{eq:IAE}
\eeq
that transform irreducibly with respect to the representation
$\bar{U}_{(12)(3)} [\Lambda ,Y]$.

The difference between the barred representation and the 
unbarred representation is that {\it the order of adding interactions and 
coupling to three-charge irreducible representation is reversed}.
In the unbarred representation interaction are added to the two-charge 
system.  The interacting two-charge system is decomposed into irreducible
representation of $ISL(2,C)$ and these are coupled to the spectator 
representation using $ISL(2,C)$ Clebsch-Gordan coefficients.  In the 
barred representation the spectator is coupled to the {\it non-interacting} 
two-charge system using the $ISL(2,C)$ Clebsch-Gordan coefficients.
The two-charge interaction $\bar{v}$ is introduced directly in this
representation.  In the absence of interactions both representation 
become equivalent.

Both set of irreducible eigenstates are complete on 
${\cal H}_{(n\pi_1)(\pi_2)}$ and the operators whose eigenvalues 
label the irreducible eigenstates in both representations have
identical spectra.  If follows that the scattering equivalence
$A_a:{\cal H}_a \to {\cal H}_a$ can be expressed as
the identity in this mixed representation:
\[
A_a = \sum \int  
\overline{\vert \vec{p}, \lambda; {q}_{(12)(3)},\vec{j},
{L}_{(12)(3)}, {S}_{(12)(3)}, \bar{m}, j_{12}}\rangle dp \times  
\]
\beq
\langle \vec{p}, \lambda; \tilde{q}_{(12)(3)},\vec{j},
\tilde{L}_{(12)(3)}, \tilde{S}_{(12)(3)}, m, j_{12}, \vert  
\label{eq:IAF}
\eeq
where the sum and integral are over the common eigenvalues of 
the corresponding observables.  The sum over the two-charge
mass eigenstates includes a bound state sum and an integral over the 
incoming or outgoing-wave scattering states in both representations.
Either choices of asymptotic condition $(\pm)$ give the same operator 
$A_a$ because the representations are scattering equivalent \cite{wp}. 
The operator $A_a$ is non-trivial if it is evaluated in a single 
representation.  By construction all of the $A_a$'s for two-cluster partitions
become the identity when interactions are turned off. 

Because the Hilbert space $H{\cal }_a$ is a proper subspace of the 
three-charge Hilbert space for each partition $a$, it is necessary to
extend the definitions  of $U_a [\Lambda,Y]$, $\bar{U}_a [\Lambda,Y]$
and $A_a$ to all of ${\cal H}$ following 
\beq
U_a [\Lambda,Y] \to 
\left (
\begin{array}{cc}
U_a [\Lambda,Y] & 0 \\
0 & I 
\end{array}
\right )
\label{eq:IAI}
\eeq
\beq
\bar{U}_a [\Lambda,Y] \to 
\left (
\begin{array}{cc}
\bar{U}_a [\Lambda,Y] & 0 \\
0 & I 
\end{array}
\right )
\label{eq:IAJ}
\eeq
\beq
A_a  \to 
\left (
\begin{array}{cc}
A_a & 0 \\
0 & I 
\end{array}
\right )
\label{eq:IAK}
\eeq

Thus the solution of the three $2+1$-charge problems gives for
each partition $a$, into at least two clusters, operators
\beq
U_a[\Lambda,Y], \qquad \bar{U}_a[\Lambda,Y],
\qquad A_a
\label{eq:IAG}
\eeq
satisfying 
\beq
\bar{U}_a[\Lambda,Y] = A_a {U}_a[\Lambda,Y] A_a^{\dagger}
\label{eq:IAH}
\eeq 
where ${U}_a[\Lambda,Y]$ is a tensor product of subsystem
representations on ${\cal H}_a$ and the mass operator $\bar{M}_a$ for
$\bar{U}_a[\Lambda,Y]$ commutes with and is independent of the
non-interacting three-charge operators, $\vec{p}$, and $h$, and
commutes with the non-interacting $j^2$.  The operators $A_a$
become the identity when the interactions are turned off.
Equations (\ref{eq:IAG}) and (\ref{eq:IAH}) also hold 
for the extended representations 
(\ref{eq:IAI},\ref{eq:IAJ}\ref{eq:IAK}) on ${\cal H}$

The representations $U_a[\Lambda,Y]$ and $\bar{U}_a[\Lambda,Y]$ are 
scattering equivalent, but only the unbarred representation satisfies 
cluster properties.

The computation of $U_a[\Lambda,Y]$, $\bar{U}_a[\Lambda,Y]$, and $A_a$ 
can all be expressed in 
terms of the solution to the mass eigenvalue problems in 
two-charge sectors. 

\section{Three-Charge Sector}
 
The construction of the dynamics in the three-charge sector
is similar to the three-particle dynamics in the 
fixed $N$ case.

The construction starts with the mass operators in the barred
representation.  The reason for introducing the barred operators,
which violate cluster properties, is that they commute with and are
independent of the total three-charge momentum and helicity of the
{\it non-interacting} system.  In addition, they commute with the
square of the spin of the non-interacting three-charge system,
independent of the partition $a$.

Mass operators $\bar{M}_a$ for each $\bar{U}_a[\Lambda ,Y]$ are 
constructed as discussed in the previous section. 
This is done for 
each of the four partitions of charges:
\beq
a=(N\pi_1)(\pi_2),(N\pi_2)(\pi_1),(\pi_1 \pi_2)(N), (N)(\pi_1)(\pi_2)
\label{eq:JA}
\eeq
These operators are easily expressed in terms of their kernels in free
three-charge irreducible bases on ${\cal H}_a$.  Note that even
thought $U_a[\Lambda ,Y]$ was extended to all of ${\cal H}$, the
generators and mass operators are non-vanishing only on the subspace
${\cal H}_a$.

A mass operator on ${\cal H}$ is defined by the linear
combination of operators using the M\"obius function of the 
lattice of partitions 
\beq
\bar{M} := \sum'_{a \in {\cal P}} C_a \bar{M}_a
+ \bar{V} \qquad C_a :=  -\mu_{1 \supset a} =
-\delta^{-1}_{1 \supset a} =  -(-1)^{n_a} (n_a-1)!
\label{eq:JB}
\eeq
where the sum is over all partitions with at least two disjoint
clusters of charge.  For two-cluster partitions $a$ of charges, like
$a=(N\pi_1)(\pi_2)$, the relation to the mass operator of the
two-charge $N\pi$ system, $\bar{M}_{N\pi}$, is of the general form
(\ref{eq:IAD}).  The combinatorial coefficients ensure the
each two-charge interaction appears only once.  The three-charge
kinetic energy on ${\cal H}_{(N)(\pi_1\pi_2)}$ appears once for each
of the three two-cluster partitions and is subtracted twice in the 
three-cluster partition.  This ensures that it appears once in the final
expression for the mass.  The M\"obius function is defined so 
this property is preserved for any number of charges and any type 
of interaction. 

The operator $\bar{V}$ is an the analog of a three-body interaction.
It vanishes when any pair of charges are separated.  In addition it
commutes with the non-interacting
three-body $j^2$ and commutes with and is independent of the
non-interacting three-body $\vec{p}$ and $h$.  In this example $\bar{V}$
includes the following types of interactions:

\begin{itemize} 

\item[1.] Three-body interaction on ${\cal H}_{N}\otimes {\cal H}_{\pi_1}
\otimes {\cal H}_{\pi_2}$.

\item[2.] $\rho-N$ interactions on ${\cal H}_{N}\otimes{\cal H}_{\rho}$.

\item[3.] $\Delta-\pi$ interactions on ${\cal H}_{\Delta_i}\otimes
{\cal H}_{\pi_j}$ for $i \not= j$.

\item[4.] Connected interactions that couple different subspaces 
in the direct sum, such as $\rho-N \leftrightarrow \pi_2-\Delta_1$
interactions, etc. 

\end{itemize} 

Since each of the operators $\bar{M}_a$, and $\bar{V}$ commute
with $j_0^2$ for the non-interacting three-charge system and commute
with and are independent of the linear momentum, $\vec{p}$, and 
helicity, $h$, for the non-interacting
three-charge system, the combined operator $\bar{M}$ also has this
property.  It follows that simultaneous eigenstates of 
$\bar{M}$, $j^2$, $\vec{p}$ and $h$ transform as mass $m$ spin $j$
irreducible representations of $ISL(2,C)$.  

The simultaneous eigenstates of $\bar{M}$, $\vec{p}_0$, $j_0^2$ and 
$h_0$ 
\beq
\vert \vec{p} , \lambda ; j,\bar{m} \rangle
\label{eq:JC}
\eeq
generally have components in all of the cluster subspaces of ${\cal H}$.  
A dynamical representation $\bar{U}[\Lambda ,Y]$ of $ISL(2,C)$ is 
given by
\beq
\bar{U}[\Lambda ,Y] \vert \vec{p} , \lambda ; j,\bar{m} \rangle =
\sum_{\lambda'=-j}^j 
\vert \vec{p}\,', \lambda' ;\bar{m},j \rangle
\sqrt{\omega_{\bar{m}} (p') \over \omega_{\bar{m}} (p)} 
D^{j}_{\lambda' \lambda} [R_H (\Lambda ,p)]e^{i p' \cdot y}. 
\label{eq:JD}
\eeq

As in the three-particle case, the eigenstates of $\bar{M}$ are obtained by
solving generalized Faddeev equations. The scattering solutions must
be solved with the appropriate asymptotic condition.  
The two-charge interactions in the three-charge Hilbert space have the from
\beq
\bar{V}_a := \sqrt{q_a^2 + (m_{120} +\bar{v})^2} -
\sqrt{q_a^2 + m_{120}^2}
\label{eq:JDA}
\eeq
and the mass operator (\ref{eq:JB}) has the form
\beq
\bar{M} = \bar{V}_{(N\pi_1)(\pi(2)} +\bar{V}_{(N\pi_2)(\pi(1)} +
\bar{V}_{(\pi_1\pi_2)(N)} + \bar{M}_{(N)(\pi_1)(\pi(2)} + \bar{V} .
\eeq

The Faddeev
equations have the same form as the corresponding non-relativistic
equations in terms of the internal kinetic energy, the interactions,
and three-charge forces.  The form of the eigenstates in a
non-interacting irreducible basis is 
\beq
\langle \vec{p}_0, \lambda_0 ; j_0 \cdots \vert
\overline{\vec{p} , \lambda ; m ,j }\rangle = 
\delta (\vec{p}_0 -\vec{p}) \delta_{\lambda_0 \lambda} \delta_{j_0 j} 
\langle \cdots \vert \overline{m} \rangle .
\label{eq:JDAA}
\eeq
The Faddeev equations (in the absence of $\bar{V}$) have the form
\beq
\vert\overline{ m} \rangle = \sum_a \vert\overline{ m;a} \rangle 
\eeq
\beq
\vert\overline{ m;a} > =  {1 \over \lambda - \bar{M}_a }\bar{V}_a \sum_{b \not= a} 
\vert \overline{m; b} \rangle
\eeq
where the indices $a,b$ correspond to two-cluster partitions.  

These equations must be solved in a fixed representation.  The
representations that are natural for the different partitions differ
by the choice of degeneracy parameters, which are dictated by the
spectator charge.  To diagonalize this operator the individual mass
operators need to be put in a common representation. This is done
using the Racah coefficients of $ISL(2,C)$ which can be computed using
four $ISL(2,C)$ Clebsch-Gordan coefficients in the same manner that
they are used to compute $SU(2)$ Racah coefficients
(\cite{wick},\cite{bkwp},
\cite{wp}).  The Racah coefficients do not depend on $\vec{p}$
or $h$, they only act on the $\cdots$ in $\langle \cdots \vert$
in (\ref{eq:JDAA}) above.   Faddeev 
equations with interactions of the general form (\ref{eq:JDA}) have been 
solved numerically for realistic interactions \cite{glockle}.   

In the representation (\ref{eq:JD}) all of the interactions are in $\bar{M}$.
In the limit that a given interaction is simply turned off we have
\beq
\bar{M} \to \bar{M}_a = A_a^{\dagger} M_a A_a
\label{eq:JE}
\eeq
which is related to the mass operator of the desired tensor product 
representation by the scattering equivalence $A_a$.  
Interactions in the operators which have domain or range 
on ${\cal H}^a$ are set to zero.  

In order to recover the desired tensor product representation 
it is enough to construct an operator $A$ with the property that 
$A \to A_a$ in the limit that the charges in different clusters 
of $a$ are asymptotically separated.   This can be done following
\cite{fcwp}\cite{wp} which use Cayley transforms:
\beq
\alpha_a := i{I- A_a  \over I+  A_1}
\label{eq:JF}
\eeq
\beq
\alpha := \alpha_{(N\pi_1)(\pi_2)} + \alpha_{(N\pi_2)(\pi_1)}+
\alpha_{(N)(\pi_1\pi_2)} 
\label{eq:JG}
\eeq  
\beq
A= {I+i \alpha \over 1- i \alpha}.
\label{eq:JH}
\eeq
The operator $A$ has the desired algebraic cluster property; which 
follows because each of the operators $\alpha_a$ vanish in the limit 
that charges in the same cluster of $a$ are separated.  The individual 
operators $A$ and $\alpha_a$ can be obtained by solving non-singular 
integral equations:
\beq
\alpha_a = i {I-A_a \over 2} + {I-A_a \over 2}\alpha_a
\label{eq:JI}  
\eeq
The operator $A$ can be obtained by solving the integral equation
\beq
B_a = {\alpha_a \over I -i \alpha_a} (I+i \alpha) + i {\alpha_a \over 
I-i \alpha_a}\sum_{b\not= a} B_b 
\label{eq:JJ}
\eeq
\beq 
A= (I + i \alpha) +i \sum_{b} B_a
\label{eq:JK}
\eeq
In the case that the $\alpha_a$'s are bounded operators 
the resulting solution is 
in the $C^*$ algebra of asymptotic constants, which means that $A$ is a 
scattering equivalence.  While the boundedness of the $\alpha_a$
have not been established in general; this property is strongly suggested by 
the structure of the expansion of the operators $A_a$ in the 
$N$-particle case
\cite{fcwp}.   

The operator
\beq
U[\Lambda ,Y] := A \bar{U}[\Lambda ,Y] A^{\dagger} 
\label{eq:JL}
\eeq
defines the desired solution of the $(2,1)$ charge sector of this model.
The dynamics is scattering equivalent to the 
$\bar{U}[\Lambda ,Y]$ dynamics and
has the property that when the interactions between charges 
in different clusters of a partition $A$ are turned off, the result is 
the tensor product of subsystem representations (on ${\cal H}_a$).  
The effect of the operators $A$ is to introduce non-trivial 
three-charge interactions into the theory. 
These interactions will not affect the spectrum or cross sections in the 
three-charge problem, but they are important contributions when the 
three-charge dynamics is used as input to a many-charge problem.
For example, they generate important exchange currents in electron 
scattering off of this three charge system, and these interactions are
also needed to imbed this model in a four-charge sector.  
Unlike the interactions $\bar{V}$, the three-charge operators generated
by cluster properties are not optional.
 
\section{Conclusion}

In this paper it was shown how to extend the construction of \cite{wp}
to construct a class of relativistic theories with cluster properties,
a spectral condition, that allow particle production.  Particle 
production requires modifications to the general construction discussed in 
\cite{wp}.  The necessary modifications were discussed in sections 
two through seven.  Rather than reprove all of the results of \cite{wp} 
using this modified framework,  the general construction was illustrated
using a non-trivial example in sections 9-11. 

Cluster properties in relativistic models is not commonly discussed,
but it is a very important topic for the experimental program at
laboratories like TJNAF.  The reason for emphasizing experiments on
few-body systems at such laboratories is that one expects that what is
learned from few-body experiments will constrain the structure of
theories that can be applied to more complex experiments.  This
requires that the many body-theories cluster to the few-body theories
that are used to the model the few-body physics.  This expectation is
trivially realized in non-relativistic quantum mechanics.  When the
reactions have sufficient energy to produce particles, a relativistic
treatment is necessary and the realization of cluster properties
becomes non-trivial.

Relativistic quantum field theory provides a formal solution to these
problems, although it is difficult to find mathematically well-defined
examples that have all of the properties that are
dictated the physical constraints.  This makes it very difficult to
find ab-initio methods to control errors in applications involving
strong interactions.  In addition, while cluster properties are
realized elegantly, there are no few-body problems; even the simplest
systems necessarily involve an infinite number of degrees of freedom.

This paper illustrates a large class of theories with all of the
desired properties.  The underlying assumption in these theories is
that number of bare-particle degrees of freedom is bounded.  This is
achieved by introducing fictitious conserved charges in the theory.
The assumption that these charges can take on only non-negative values
and each particle has at least one positive charge limits the number
of degrees of freedom.  If these conditions are relaxed, the resulting
theory will involve an infinite number of degrees of freedom.
 
The theories constructed in this paper have meaningful few-body
problems.  In the three-charge model it was shown that the two-charge
models determine the interactions in the charge sector up to a
three-charge interaction.  In general, $K$-charge interactions in the
$N$-charge problem are determined from the $K$-charge problem using
cluster properties.  Like the fixed particle number case, cluster
properties introduce non-trivial many-charge interactions into the
dynamics.  These interactions are determined recursively by the fewer-charge
interactions in the absence of an explicit $N$-charge interaction like
$\bar{V}$.  In general, just like with the fixed number of particle
case, the few and many-charge interactions mix under change of
representation.  The construction in this paper can be used to
formulate relativistic isobar models and models with a dynamics
dominated by resonances.

It is desirable to go beyond the restrictions imposed by charge
conservation.  The requirement of having a meaningful few-body problem
puts strong constraints on how cluster properties should be
implemented in the general theory.  One way to control the number of
degrees of freedom and have a meaningful few-body problem is to
reformulate the theory so the relevant degrees of freedom are
physical-particle degrees of freedom.  In this way the center of
momentum energy controls the number of degrees of freedom.  In this
picture physical particles play the same role as minimally charged
particles.  The mechanics of coupling the physics on different energy 
scales provides the interesting challenge which needs to be 
addressed to extend the construction of this paper.
 
The models discussed in this paper are valuable precisely because they
are quantum models with an exact Poincar\'e symmetry which also
satisfies cluster properties.  In the absence of a more fundamental
theory, cluster properties and experiments on subsystems put strong
constraints on the relativistic many-charge dynamics, which can then
be used to make predictions of the theory.

\end{document}